\documentclass[aps,prb,twocolumn,floatfix,groupedaddress]{revtex4-2}
\usepackage{graphicx,amsmath,amsfonts,mathrsfs,bbold,dsfont}
\usepackage{xcolor}
\begin{document}

\title{Hidden ferromagnetism of centrosymmetric antiferromagnets}

\author{I.~V.~Solovyev}
\email{SOLOVYEV.Igor@nims.go.jp}
\affiliation{Research Center for Materials Nanoarchitectonics (MANA), National Institute for Materials Science (NIMS), 1-1 Namiki, Tsukuba, Ibaraki 305-0044, Japan}

\date{\today}

\date{\today}
\begin{abstract}
The time-reversal symmetry ($\mathcal{T}$) breaking is a signature of ferromagnetism, giving rise to such phenomena as the anomalous Hall effect (AHE) and orbital magnetism. Nevertheless, $\mathcal{T}$ can be also broken in certain classes of antiferromagnets, such as weak ferromagnets or altermagnets, which remain invariant under the spatial inversion. In the light of this similarity with the ferromagnetism, it is tempting to ask whether such unconventional antiferromagnetic (AFM) state can be represented as the simplest ferromagnetic one, i.e. within the minimal unit cell containing only one magnetic site. We show that such representation is possible due to special form of the spin-orbit (SO) interaction in an antipolar lattice hosting this AFM state. The inversion symmetry constrains the form of the SO interaction, which becomes invariant under the symmetry operation $\{ \mathcal{S}| {\bf t} \}$, combining the $180^{\circ}$ rotation of spins ($\mathcal{S}$) with the lattice shift ${\bf t}$, connecting two antiferromagnetically coupled sublattices. This is the fundamental symmetry property of centrosymmetric antiferromagnets, which justifies the use of the generalized Bloch theorem and transformation to the local coordinate frame with one magnetic site per cell. It naturally explains the emergence of AHE and net orbital magnetization, and provide transparent expressions for these properties in terms of the electron hoppings and SO interaction operating between AFM sublattices, as well as the orthorhombic strain, controlling the piezomagnetic response. The idea is illustrated on a number of examples including two-dimensional square lattice, monoclinic VF$_4$ and CuF$_2$, and RuO$_2$-type materials with the rutile structure, using for these purposes realistic models derived from first-principles calculations. 
\end{abstract}

\maketitle

\section{\label{sec:Intro} Introduction}
\par Canonically, the time-reversal symmetry ($\mathcal{T}$) breaking was synonymous with ferromagnetism. The magnetic unit cell of a regular ferromagnet coincides with the chemical one, but the electronic structure for the states with spins ``up'' and ``down'' is different, so that the ferromagnetic (FM) system is characterized by a finite spin magnetic moment, which has the same direction at all magnetic sites. On the contrary, the regular antiferromagnet is characterised by the doubling of the unit cell, so that two magnetic sublattices, which in the FM state would be connected by a primitive translation ${\bf t}$, are occupied by atoms with opposite directions of spins. The basic symmetry of the regular antiferromagnetic (AFM) state is $\{ \mathcal{T}|{\bf t} \}$, where $\mathcal{T}$ is combined with the lattice translation ${\bf t}$ of the chemical cell. Therefore, although $\mathcal{T}$ is microscopically broken, it is preserved macroscopically, after averaging over two AFM sublattices~\cite{Dzyaloshinskii1991}. The electronic states of such antiferromagnets are Kramers degenerate. 

\par Currently, a great deal of attention is paid to the systems, where the AFM alignment of spins coexists with some features, which are more characteristic for ferromagnets, including the macroscopic time-reversal symmetry breaking and splitting of AFM bands. Such systems are now called ``altermagnets'', to emphasize the distinct character of these materials and their principal difference from the conventional ferromagnets and antiferromagnets~\cite{SmejkalPRX1,SmejkalPRX2,MazinPRX1}. 

\par On the other hand, the AFM materials with broken time-reversal symmetry have been known for decades. Particularly, in 1950s, using phenomenological symmetry arguments, Dzyaloshinskii pointed out that there are special types of antiferromagnets, which can host the phenomena of weak ferromagnetism~\cite{Dzyaloshinskii_weakF}, piezomagnetism~\cite{DzyaloshinskiiPM}, and magnetoelectricity~\cite{DzyaloshinskiiME}. Contrary to the regular antiferromagnetism, the magnetic unit cell of these material coincides with the chemical one, being the necessary precondition for macroscopic time-reversal symmetry breaking~\cite{Dzyaloshinskii1991}. A very detailed phenomenological classification has been given by Turov, who suggested to divide unconventional antiferromagnets into two major classes, calling them ``centrosymmetric'' and ``centroantisymmetric'', depending on whether the spatial inversion $\mathcal{I}$ enters the magnetic group alone or in the combination with  $\mathcal{T}$~\cite{TurovBook}. The centrosymmetric antiferromagnetism in Turov's classification and encompassing such phenomena as weak ferromagnetism and piezomagnetism has a clear resemblance to what is now called altermagnetism, where the AFM alignment of spins can induce a spontaneous magnetization and other ferromagentic properties. The centroantisymmetric antiferromagnetism provides a general framework for understanding the magnetoelectricity, which excludes any ferromagnetic behavior in the ground state unless an external magnetic or electric field is applied. Besides the weak ferromagnetism, piezomagnetism, and magnetoelectricity, Turov consider a wide spectrum of phenomena expected in the unconventional antiferromagnets. Particularly, in 1962, Turov and Shavrov have predicted the anomalous Hall effect (AHE) in centrosymmetric antiferromagnets. Moreover, they have explicitly stated that AHE is not a side effect of the weak ferromagnetism. Rather, it can be related directly to the AFM order parameter~\cite{TurovBook,TurovShavrov}. The Landau theory of altermagnetism, which was recently formulated in Ref.~\cite{McClartyRau} and applied to rutile and hexagonal (MnTe) materials, successfully reproduces the behavior of net magnetization and AHE predicted in Ref.~\cite{TurovBook}. Similar symmetry-based arguments can be found in recent review article~\cite{SmejkalNRM}. Turov also pointed out that since magnetoelectricity and weak ferromagnetism belong to two different classes, they are mutually exclusive and do not coexist unless they develop in different magnetic sublattices~\cite{TurovUFN}. The typical example of such coexistence is GdFeO$_3$, where Gd sublattice is magnetoelectric, while Fe sublattice is weakly ferromagnetic~\cite{Yamaguchi,Tokunaga}.

\par Although many properties of centrosymmetric antiferromagnets were anticipated on the phenomenological level, recent breakthrough in this field is related to microscopic understanding of these properties, which became largely possible due to development of first-principles electronic structure calculations. The main attention is paid to the search of new materials with large spin splitting of AFM bands~\cite{Okugawa,HayamiJPSJ,Naka,SmejkalSA,NakaOrganic,JunweiLiuNC,Naka2022,OghushiMnTe,BelashchenkoMnTe,JunweiLiuPRX}, which is regarded as a hallmark of altermagnetism~\cite{SmejkalPRX1,SmejkalPRX2,MazinPRX1,LingBai}. 

\par Such splitting is a consequence of the rotational symmetry of crystals with several atoms in the unit cell. For nonsymmorphic symmetries, the atomic positions are generated by two- or fourfold rotations $\mathcal{C}$, which are combined with the with the lattice shift ${\bf t}$ connecting different sublattices: $\{ \mathcal{C}|{\bf t} \}$. The symmorphic analogs, containing fourfold rotations alone, are also possible, but less common~\cite{SmejkalPRX1,LingBai}. Then, the AFM order is obtained by combining some of the symmetry operations $\{ \mathcal{C}|{\bf t} \}$ with $\mathcal{T}$, so that the magnetic unit cell remains the same as the chemical one, being in line with Dzyaloshinskii's conjecture~\cite{Dzyaloshinskii1991}. The new aspect of the problem, which was overlooked in the earlier stages, is that the $\{ \mathcal{T} \mathcal{C}|{\bf t} \}$ symmetry gives rise to the spin-splitting of bands in the reciprocal space~\cite{SmejkalPRX1,SmejkalPRX2,LingBai,Naka_Spintronics}. Today, this splitting is believed to be the sole signature of ferromagnetism in otherwise AFM materials, which is responsible for AHE and other ferromagnetic phenomena.

\par However, there is another important symmetry, $\mathcal{I}$, which was highlighted in Turov's definition of ``centrosymmetric antiferromagnetism'', but somewhat overshadowed by other symmetry properties in the modern developments of altermagnetism. Then, which symmetry is more important, for instance for the emergence of AHE in AFM substances: $\{ \mathcal{T} \mathcal{C}|{\bf t} \}$ or $\mathcal{I}$?

\par In our recent work, dealing with the minimal one-orbital model, where the spin-orbit (SO) interaction replicates the form of Dzyaloshinskii-Moriya (DM) interactions in the noncentrosymmetric bonds~\cite{Dzyaloshinskii_weakF,Moriya_weakF}, we have argues that the altermagnetic splitting of bands does not play a key role in AHE and orbital magnetism~\cite{arXiv2025}. Namely, $\mathcal{T}$ can be broken even when the AFM bands are spin-degenerate. Similar behaviour has been found in a more sophisticated multi-orbital model for AFM $\kappa$-type organic conductors and orthorhombically distorted perovskites~\cite{NakaOrganic,Naka2022}. Furthermore, we have argued that the fundamental symmetry of such AFM state is $\{ \mathcal{S}| {\bf t} \}$, which combines the $180^{\circ}$ rotation of spins $\mathcal{S} = i\hat{\sigma}_{y}$ ($\hat{\sigma}_{y}$ being the Pauli matrix) with the lattice shift ${\bf t}$, connecting antiferromagnetically coupled sublattices. This means that the eigenvectors for the spin states $\uparrow$ and $\downarrow$ differ only by a phase factor, which guarantees that (i) the spin bands are degenerate and (ii) the contributions of these bands to the anomalous Hall conductivity are equal to each other and, instead of the partial cancellation, which would occur in regular ferromagnets, we have an \emph{addition} of such contributions~\cite{arXiv2025}. The situation is fundamentally different from the $\mathcal{IT}$ symmetry realized in magnetoelectric materials~\cite{DzyaloshinskiiME,TurovBook,TurovUFN}, where the bands are also degenerate, but the contributions of these bands to AHE cancel each other~\cite{ChenNiuMacDonald}.

\par The time-reversal operation is the combination of $\mathcal{S}$ and the complex conjugation $K$: $\mathcal{T}=\mathcal{S}K$. Therefore, if the microscopic Hamiltonian is complex, the symmetry operation $\{ \mathcal{S}| {\bf t} \}$ is not the same as $\{ \mathcal{T}| {\bf t} \}$, which is expected for regular antiferromagnets. That is why the spin degeneracy can exist even when the time-reversal symmetry is broken. The spin degeneracy in this case is \emph{not} the Kramers' degeneracy, because the latter implies that the system is $\mathcal{T}$-invariant, which is obviously not the case here.

\par In this work we will explicitly show that the $\{ \mathcal{S}| {\bf t} \}$ symmetry is directly related to the fact that $\mathcal{I}$ is conserved in the centrosymmetric antiferromagnets, which imposes a severe constraint on the form of the SO interaction in magnetic bonds. It is true that $\mathcal{I}$ transforms each sublattice to itself and does not interconnect the sites belonging to different sublattices~\cite{footnote}. However, $\mathcal{I}$ will unambiguously specify the properties of magnetic bonds: namely, knowing the bond properties around one magnetic site, one can find the properties around neighboring site, belonging to different sublattice. Particularly, we will explicitly show that the SO interaction on the centrosymmetric antipolar lattice behaves as an AFM object and changes its sign when it moves from one sublattice to another. That is why it collaborates with the AFM N\'eel field and leads to the $\{ \mathcal{S}| {\bf t} \}$ symmetry of microscopic Hamiltonian.

\par There is another important aspect of the $\{ \mathcal{S}| {\bf t} \}$ symmetry: the spin rotation $\mathcal{S}$ is combined with the shift ${\bf t}$, which can be viewed as the lattice translation of a more compact unit cell with only one magnetic site. This constitutes the basis of the so-called generalized Bloch theorem~\cite{Sandratskii_review}, which states that by a unitary transformation to the local coordinate frame, where all the spins are pointed in the positive direction of $z$, the AFM system can be described within this compact unit cell as if it would be ``ferromagnetic'' one. Thus, the generalized Bloch theorem provides a mapping of the AFM system onto a FM one, which naturally explains the breaking of the time-reversal symmetry and emergence of associated with it magnetic effect, which are originally known for ferromagnets, but can be realized in some AFM systems.  

\par The rest of the article is organized as follows. In Sec.~\ref{sec:basic} we will consider the basic symmetry properties of SO interaction imposed by $\mathcal{I}$ in the combination with lattice translations. Then, in Sec.~\ref{sec:AFE}, we will show that the behavior of the SO interactions is related to even more fundamental properties of the magnetoelectric coupling in antipolar lattices. Particularly, in order to preserve $\mathcal{I}$, the distortion must be antipolar, which inevitably leads to the doubling of the unit cell and the sign change of the SO interaction when it is considered around two sublattices. Then, in Sec.~\ref{sec:gBloch}, we will apply the generalized Bloch theorem and show that, for some components of the SO interaction, the sign change can be compensated by the transformation to the local coordinate frame, so that the AFM system can be formally described as a FM one, with smaller unit cell containing only one magnetic site. In Sec.~\ref{sec:mmodel}, we will introduce the minimal model for the centrosymmetric antiferromagnets with broken $\mathcal{T}$ and derive transparent expressions for the anomalous Hall conductivity and orbital magnetization in the local coordinate frame. Applications for the square lattice, monoclinic VF$_4$ and CuF$_2$, as well as RuO$_2$-type materials with the tetragonal symmetry will be considered in Sec.~\ref{sec:examples}. In Sec.~\ref{sec:separation} we will explain how different contribution to AHE, associated with spin-degenerate and nondegenerate bands, can be evaluated in realistic electronic structure calculations. Finally, in Sec.~\ref{sec:Summary}, we will summarize our work. Two appendices deal with the unitary transformation of the SO interaction and elimination of the same-sign components responsible for the weak spin ferromagnetism, and construction of the model Hamiltonian for VF$_4$ and CuF$_2$ on the basis of first-principles electronic structure calculations.

\section{\label{sec:basic} Basic considerations}
\par Let ${\bf T}_{1}$, ${\bf T}_{2}$, and ${\bf T}_{3}$ be the primitive translations of a crystal, consisting of two sublattices, which can be transformed to each other by a rotation, combined with the shift of the origin by ${\bf t} = \frac{1}{2} \left( {\bf T}_{1} + {\bf T}_{2} + {\bf T}_{3} \right)$. Therefore, if one sublattice is centered at the origin ${\bf 0} = (0,0,0)$, another sublattice will be centered at ${\bf t}$, and if ${\bf 0}$ is an inversion center, ${\bf t}$ is another inversion center~\cite{footnote}. Nevertheless, the midpoint between ${\bf 0}$ and ${\bf t}$ is not the inversion center, so that there is a finite DM interactions operating between different sublattices~\cite{Dzyaloshinskii_weakF,Moriya_weakF}. 

\par The atomic positions in both the sublattices can be generated by the vectors ${\bf t}_{1}=\frac{1}{2}(-{\bf T}_{1} + {\bf T}_{2} + {\bf T}_{3})$, ${\bf t}_{2}=\frac{1}{2}({\bf T}_{1} - {\bf T}_{2} + {\bf T}_{3})$, and ${\bf t}_{3}=\frac{1}{2}({\bf T}_{1} + {\bf T}_{2} - {\bf T}_{3})$. However, ${\bf t}_{1}$, ${\bf t}_{2}$, and ${\bf t}_{3}$ do not necessarily transform the DM interactions to themselves and in this sense are not the primitive translations. In any case, the atomic positions are fully specified by the vectors $\boldsymbol{R} = l {\bf t}_{1} + m {\bf t}_{2} + n {\bf t}_{3}$. If $l+m+n$ is even, the atom belong to the sublattice $1$. If it is odd, the atom belongs to the sublattice $2$. Thus, the DM interactions, $\boldsymbol{D}_{\boldsymbol{R},\boldsymbol{R}'}$, operating between different sublattices, are finite if $\boldsymbol{R}' - \boldsymbol{R}$ is an odd superposition of ${\bf t}_{1}$, ${\bf t}_{2}$, and ${\bf t}_{3}$. 

\par The DM interactions are induced by the SO coupling. In the simplest one-orbital model, the SO interactions are given by $\hat{\cal H}^{\rm so}_{\boldsymbol{R},\boldsymbol{R}'} = i \boldsymbol{t}_{\boldsymbol{R},\boldsymbol{R}'} \cdot \hat{\boldsymbol{\sigma}}$, where $\hat{\boldsymbol{\sigma}}$ is the vector of Pauli matrices and $\boldsymbol{D}_{\boldsymbol{R},\boldsymbol{R}'} \sim \boldsymbol{t}_{\boldsymbol{R},\boldsymbol{R}'}$~\cite{Shekhtman,review2024}. Therefore, $\boldsymbol{t}_{\boldsymbol{R},\boldsymbol{R}'}$ and $\boldsymbol{D}_{\boldsymbol{R},\boldsymbol{R}'}$ obey the same symmetry rules.

\par
\begin{figure}[b]
\begin{center}
\includegraphics[width=8.2cm]{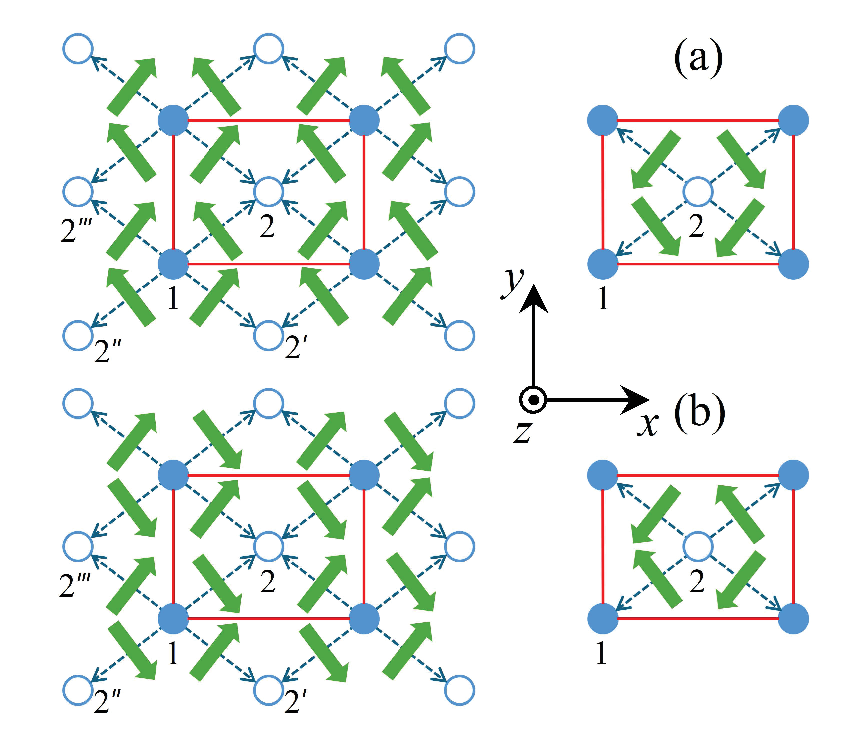} 
\end{center}
\caption{Parameters of SO interaction around the atoms of two sublattices in the centrosymmetric structure obeying the $\{ {\cal C}_{2x} | {\bf t} \}$ symmetry (a, top) and $\{ {\cal C}_{2y} | {\bf t} \}$ symmetry (b, bottom). The atoms of the sublattices $1$ and $2$ are shown by filled and open circles, respectively. The bond directions, which are defined starting from the central site in the direction of neighboring sites, are shown by broken arrows. The vectors $\boldsymbol{t}_{\boldsymbol{R},\boldsymbol{R}'}$, attached to these bonds, are shown by the bold green arrows. According to this definition, the bonds around the sites $1$ and $2$ have opposite directions, which also flip the directions of the vectors $\boldsymbol{t}_{\boldsymbol{R},\boldsymbol{R}'}$ attached to these bonds, as shown on the left and right parts of the figure. The unit cell is shown by the solid red line.}
\label{fig:DM}
\end{figure}

 The combination of translational invariance and ${\cal I}$ impose a severe constraint on the form of $\hat{\cal H}^{\rm so}_{\boldsymbol{R},\boldsymbol{R}'}$. ${\cal I}$ requires that for each bond $\boldsymbol{R}' - \boldsymbol{R}$, there should be the bond $\boldsymbol{R}'' - \boldsymbol{R}$ in the opposite direction, as explained in Fig.~\ref{fig:DM}, where $1$-$2''$ is the bond opposite to $1$-$2$ and $1$-$2'''$ is the bond opposite to $1$-$2'$. 
\noindent
\noindent Then, ${\cal I}$ does not change the axial vector $\boldsymbol{t}_{\boldsymbol{R},\boldsymbol{R}'}$ and, therefore, $\boldsymbol{t}_{\boldsymbol{R},\boldsymbol{R}'} = \boldsymbol{t}_{\boldsymbol{R},\boldsymbol{R}''}$. Furthermore, each atomic site is surrounded by the atoms of another sublattice. The atoms in one sublattice are translated to each other by ${\bf T}_{1}$, ${\bf T}_{2}$, and ${\bf T}_{3}$, which also translate the vectors $\boldsymbol{t}_{\boldsymbol{R},\boldsymbol{R}'}$ around these atoms as shown in Fig.~\ref{fig:DM}. Then, since $\hat{\cal H}^{\rm so}_{\boldsymbol{R},\boldsymbol{R}'}$ is the hermitian matrix, $\boldsymbol{t}_{\boldsymbol{R},\boldsymbol{R}'}$ should satisfy the property $\boldsymbol{t}_{\boldsymbol{R}',\boldsymbol{R}} = - \boldsymbol{t}_{\boldsymbol{R},\boldsymbol{R}'}$. Therefore, if we take $\boldsymbol{R}$ as a central site and define the directions of the bond starting from this $\boldsymbol{R}$ and going to the neighboring site $\boldsymbol{R}'$, the parameters $\boldsymbol{t}_{\boldsymbol{R}',\boldsymbol{R}}$ around the site $\boldsymbol{R}'$ belonging to another sublattice will differ from those around $\boldsymbol{R}$ by the sign, as is clearly seen in Fig.~\ref{fig:DM}. Thus, for an arbitrary $\boldsymbol{R}'' = l'' {\bf t}_{1} + m'' {\bf t}_{2} + n'' {\bf t}_{3}$, ${\cal I}$ imposes the following constraint:
\noindent
\begin{equation}
\boldsymbol{t}_{\boldsymbol{R}+\boldsymbol{R}'',\boldsymbol{R}'+\boldsymbol{R}''} = (-1)^{l''+m''+n''} \, \boldsymbol{t}_{\boldsymbol{R},\boldsymbol{R}'}.
\label{eq:tHso}
\end{equation}

\par The remaining uncertainty is how the parameters $\boldsymbol{t}_{\boldsymbol{R},\boldsymbol{R}'}$ behave in the bonds, which are not connected by ${\cal I}$, such as the bonds $1$-$2$ and $1$-$2'$ in Fig.~\ref{fig:DM}. This depends on other symmetries. For instance, in the two-dimensional case, the sublattices can be connected combining the twofold rotations about either $x$ or $y$ with ${\bf t}$: $\{ {\cal C}_{2x} | {\bf t} \}$ and $\{ {\cal C}_{2y} | {\bf t} \}$, respectively. The symmetry operation $\{ {\cal C}_{2x} | {\bf t} \}$ ($\{ {\cal C}_{2y} | {\bf t} \}$) will not only change the signs of the $y$ ($x$) and $z$ components of $\boldsymbol{t}_{\boldsymbol{R},\boldsymbol{R}'}$, but also interchange the sublattices. Therefore, around each site, the $x$ ($y$) component of $\boldsymbol{t}_{\boldsymbol{R},\boldsymbol{R}'}$ will be sign-alternating, while two other components will have the same sign in all the bonds. The corresponding patters of $\boldsymbol{t}_{\boldsymbol{R},\boldsymbol{R}'}$ are shown in Figs.~\ref{fig:DM}(a) and (b). The sign-alternating component is responsible for AHE and orbital magnetism~\cite{NakaOrganic,arXiv2025}, while the same-sign components give rise to the weak spin ferromagnetism~\cite{Dzyaloshinskii_weakF,Shekhtman}. Thus, it is clear that weak spin ferromagnetism and AHE have different microscopic origin and the latter cannot be described in terms of net spin magnetization~\cite{SmejkalSA,ChenNiuMacDonald,PRB1997}. 

\section{\label{sec:AFE} Antipolar Distortions and Weak Ferromagnetism}
\par In this section, we will elucidate even more fundamental reason why the DM interactions for the weak ferromagnets have a specific pattern shown in Fig.~\ref{fig:DM}, which preserves the inversion symmetry, but doubles the unit cell, resulting in two sublattices. We believe that the most fundamental quantity to consider is magnetoelectric coupling $\vec{\boldsymbol{\mathcal{P}}}_{\boldsymbol{R},\boldsymbol{R}'}$, relating the cross product of spins in the bond to the electric polarization in the same bond, $\vec{P}_{\boldsymbol{R},\boldsymbol{R}'} = \vec{\boldsymbol{\mathcal{P}}}_{\boldsymbol{R},\boldsymbol{R}'} \cdot [\boldsymbol{e}_{\boldsymbol{R}} \times \boldsymbol{e}_{\boldsymbol{R}'} ]$, where $\boldsymbol{e}_{\boldsymbol{R}}$ is the unit vector in the direction of spin at the site $\boldsymbol{R}$~\cite{KNB,PRL2021,MDPI2025}. 

\par $\vec{\boldsymbol{\mathcal{P}}}_{\boldsymbol{R},\boldsymbol{R}'} \equiv [\mathcal{P}_{\boldsymbol{R},\boldsymbol{R}'}^{v,c}]$ is the $3 \times 3$ tensor. According to our conventions~\cite{MDPI2025}, the bold character refers to the spin components ($c$), which couple to the cross product, while the vector symbol refers to the $x$, $y$, and $z$ components ($v$) of electric polarization. Thus, under the spatial inversion, the $v$ components will transform according to the normal vector rules, while the $c$ components will transform according to the pseudovector rule. 

\par $\vec{\boldsymbol{\mathcal{P}}}_{\boldsymbol{R},\boldsymbol{R}'}$ can be finite in centrosymmetric bonds. Indeed, the spatial inversion about the center of the bond yields the following property: ${\cal I}\vec{\boldsymbol{\mathcal{P}}}_{\boldsymbol{R},\boldsymbol{R}'} = - \vec{\boldsymbol{\mathcal{P}}}_{\boldsymbol{R}',\boldsymbol{R}} = \vec{\boldsymbol{\mathcal{P}}}_{\boldsymbol{R},\boldsymbol{R}'}$, which means that $\vec{\boldsymbol{\mathcal{P}}}_{\boldsymbol{R},\boldsymbol{R}'}$ does not necessarily vanish. This is the reason why the electric polarization $\vec{P}$ in insulators can be induced by a noncollinear alignment of spins in otherwise centrosymmetric crystals~\cite{KNB,PRL2021}. Alternatively, the electric field $\vec{E}_{\boldsymbol{R},\boldsymbol{R}'}$, acting in the bond, will interact with $\vec{\boldsymbol{\mathcal{P}}}_{\boldsymbol{R},\boldsymbol{R}'}$ and makes the spins $\boldsymbol{e}_{\boldsymbol{R}}$ and $\boldsymbol{e}_{\boldsymbol{R}'}$ noncollinear. The corresponding interaction energy is given by $\boldsymbol{D}_{\boldsymbol{R},\boldsymbol{R}'} \cdot [\boldsymbol{e}_{\boldsymbol{R}} \times \boldsymbol{e}_{\boldsymbol{R}'} ]$, where $\boldsymbol{D}_{\boldsymbol{R},\boldsymbol{R}'} = - \vec{E}_{\boldsymbol{R},\boldsymbol{R}'} \cdot \vec{\boldsymbol{\mathcal{P}}}_{\boldsymbol{R},\boldsymbol{R}'}$ is the DM interaction induced by the electric field. 

\par For insulators, $\vec{\boldsymbol{\mathcal{P}}}_{\boldsymbol{R},\boldsymbol{R}'}$ can be evaluated in terms of the superexchange theory, similar to the DM coupling. Indeed, the latter is given by $\boldsymbol{D}_{\boldsymbol{R},\boldsymbol{R}'} = \frac{2t_{\boldsymbol{R},\boldsymbol{R}'}\boldsymbol{t}_{\boldsymbol{R},\boldsymbol{R}'}}{U}$, while the former is $\vec{\boldsymbol{\mathcal{P}}}_{\boldsymbol{R},\boldsymbol{R}'} = -\frac{2e}{V} \frac{t_{\boldsymbol{R},\boldsymbol{R}'} \vec{\boldsymbol{r}}_{\boldsymbol{R},\boldsymbol{R}'}}{U}$, where $-e$ is the electron charge, $V$ is the unit cell volume, $U$ is the on-site Coulomb repulsion, $\hat{t}_{\boldsymbol{R},\boldsymbol{R}'} = t_{\boldsymbol{R},\boldsymbol{R}'}\hat{\mathbb{1}} + i\boldsymbol{t}_{\boldsymbol{R},\boldsymbol{R}'} \cdot \hat{\boldsymbol{\sigma}}$ is the transfer integral expanded in terms of the $2 \times 2$ unity matrix $\hat{\mathbb{1}}$ and the vector of Pauli matrices, and $\hat{\vec{r}}_{\boldsymbol{R},\boldsymbol{R}'} = \vec{r}_{\boldsymbol{R},\boldsymbol{R}'}\hat{\mathbb{1}} + i\vec{\boldsymbol{r}}_{\boldsymbol{R},\boldsymbol{R}'} \cdot \hat{\boldsymbol{\sigma}}$ is similar expansion for the position operator~\cite{PRL2021}.

\par
\begin{figure}[b]
\begin{center}
\includegraphics[width=8.2cm]{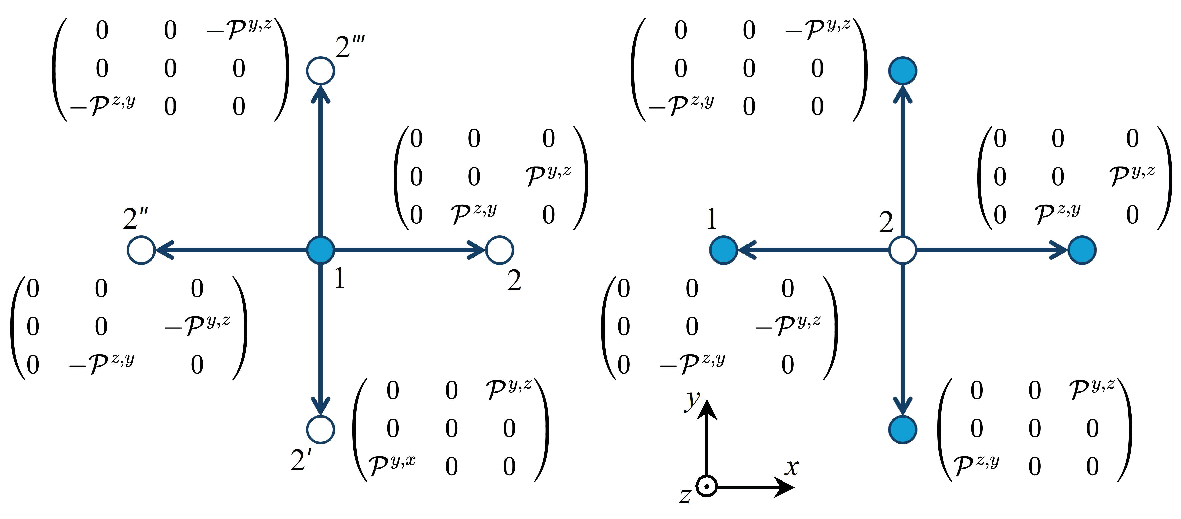} 
\end{center}
\caption{The form of magnetoelectric tensors $\vec{\boldsymbol{\mathcal{P}}}_{\boldsymbol{R},\boldsymbol{R}'}$ in the bonds around two neighboring sites in the ideal square lattice. The directions of the bonds are indicated by arrows.}
\label{fig:P}
\end{figure}

 $\vec{\boldsymbol{\mathcal{P}}}_{\boldsymbol{R},\boldsymbol{R}'}$ obeys certain symmetry properties. To be specific, let us consider the ideal square lattice, which can be transformed to itself by the fourfold rotations around $z$ (${\cal C}_{4z}$) and the spatial inversion about the lattice sites. The primitive translations are ${\bf t}_{1}$ and ${\bf t}_{2}$, so that each lattice point is specified by $\boldsymbol{R} = m \, {\bf t}_{1} + n \, {\bf t}_{1}$. Furthermore, each bond obeys the twofold rotational symmetry about itself. Then, for the single bond along $x$, say $1$-$2$ in Fig.~\ref{fig:P}, this tensor will have the following form~\cite{MDPI2025}:
\noindent
\begin{displaymath}
\vec{\boldsymbol{\mathcal{P}}}_{1,2} =
\left(
\begin{array}{ccc}
\mathcal{P}^{x,x} & 0                 & 0                 \\
0                 & \mathcal{P}^{y,y} & \mathcal{P}^{y,z} \\
0                 & \mathcal{P}^{z,x} & \mathcal{P}^{z,z} 
\end{array}
\right).
\end{displaymath}
\noindent The bond $1$-$2$ can be transformed to the bond $1$-$2''$ by ${\cal I}$, which changes the sign of the whole magnetoelectric tensor, $\vec{\boldsymbol{\mathcal{P}}}_{1,2''} = - \vec{\boldsymbol{\mathcal{P}}}_{1,2}$. Alternatively, $1$-$2$ can be transformed to $1$-$2''$ by the twofold rotation about $z$, which changes the sign of only off-diagonal elements. The combination of these symmetry properties imposes a constraint on the form of $\vec{\boldsymbol{\mathcal{P}}}_{\boldsymbol{R},\boldsymbol{R}'}$, which is summarized in Fig.~\ref{fig:P}: for each bond, the tensor $\vec{\boldsymbol{\mathcal{P}}}_{\boldsymbol{R},\boldsymbol{R}'}$ is specified by only two elements, $\mathcal{P}^{y,z}$ and $\mathcal{P}^{z,y}$. The Katsura–Nagaosa–Balatsky rule~\cite{KNB}, which can be justified for high symmetries of the bond~\cite{MDPI2025}, additionally requires $\mathcal{P}^{z,y} = -\mathcal{P}^{y,z}$~\cite{KNB,PRL2021}. However, for lower symmetries this rule does not apply~\cite{MDPI2025}, so that generally $\mathcal{P}^{z,y}$ and $\mathcal{P}^{y,z}$ are two independent parameters. Nevertheless, such details are not important for our consideration. 
\noindent

\par The magnetoelectric tensor is periodic and, therefore, $\vec{\boldsymbol{\mathcal{P}}}_{\boldsymbol{R}+\boldsymbol{R}'',\boldsymbol{R}'+\boldsymbol{R}''} = \vec{\boldsymbol{\mathcal{P}}}_{\boldsymbol{R},\boldsymbol{R}'}$ for any $\boldsymbol{R}''$ (for instance, site $1$ in Fig.~\ref{fig:P} can be moved to site $2$, etc.). 

\par Next, let us consider possible patters of electric fields caused by internal atomic displacements. For simplicity, we assume that all fields are parallel to $z$, but can have different directions in different bonds. For instance, such fields can be due to the buckling of the TM-L-TM bonds, where the intermediate ligand (L) sites are displaced parallel to $z$ relative to the transition-metal (TM) sites. Then, there are three main patterns, which are explained in Fig.~\ref{fig:E}: 
\noindent
\begin{figure}[t]
\begin{center}
\includegraphics[width=8.2cm]{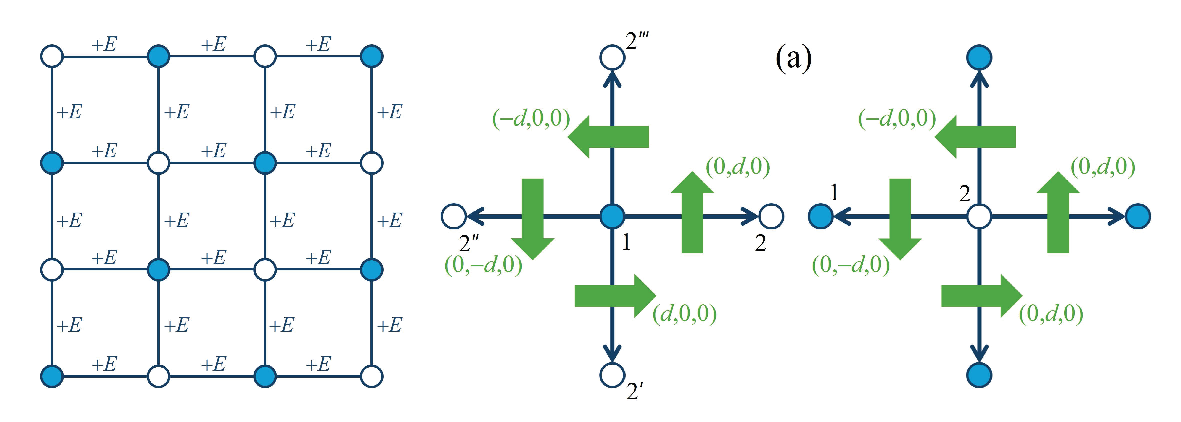} 
\end{center}
\begin{center}
\includegraphics[width=8.2cm]{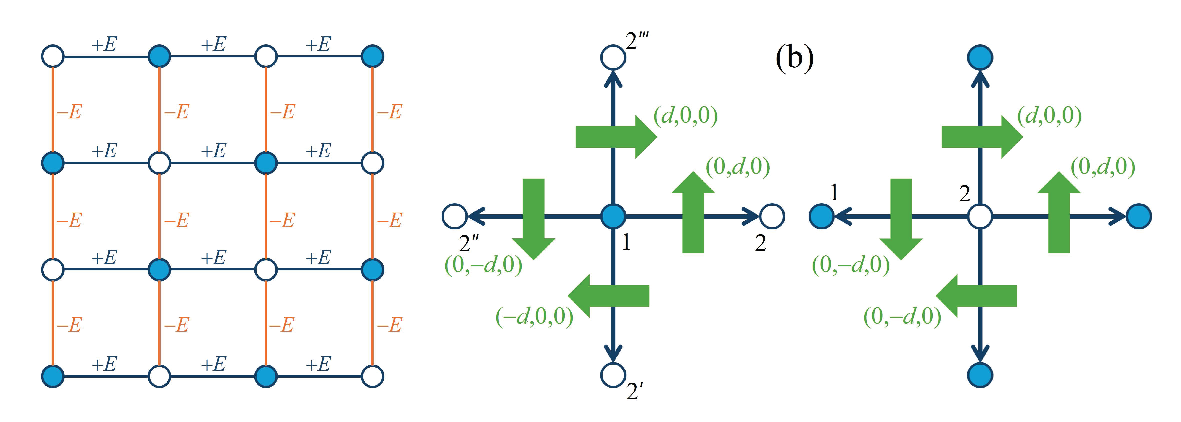} 
\end{center}
\begin{center}
\includegraphics[width=8.2cm]{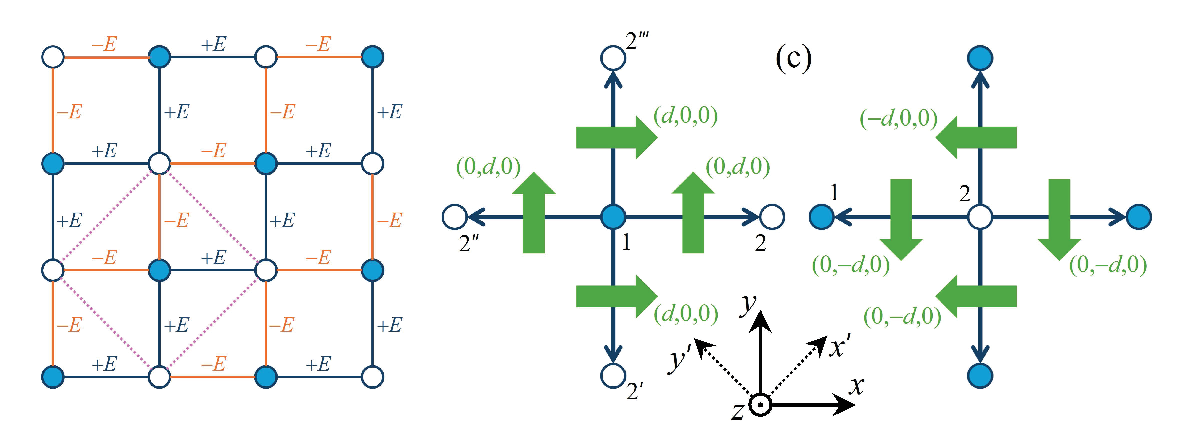} 
\end{center}
\caption{Main patters of internal electric fields (left) and corresponding to them Dzyaloshinskii-Moriya interactions (right) on the square lattice: (a) polar, (b) antipolar noncentrosymmetric, and (c) antipolar centrosymmetric. The doubled unit cell in the latter case is shown by broken line. $xy$ is the coordinate frame of the regular square lattice. $x'y'$ is the coordinate frame, which is typically used for the doubled unit cell, such as in Fig.~\ref{fig:DM}.}
\label{fig:E}
\end{figure}
\noindent
\begin{itemize}
\item[(a)] Polar pattern, where all $\vec{E}_{\boldsymbol{R},\boldsymbol{R}'}$ are in the positive direction of $z$. Since $\vec{E}$ is the normal vector, the inversional invariance would requite $\vec{E}_{\boldsymbol{R},\boldsymbol{R}'} = -\vec{E}_{\boldsymbol{R},\boldsymbol{R}''}$ for each $\boldsymbol{R}'$ and $\boldsymbol{R}''$ located on the opposite sides of the inversion center $\boldsymbol{R}$. Therefore, the polar pattern breaks ${\cal I}$. Nevertheless, the displacements are periodic and $\vec{E}_{\boldsymbol{R}+\boldsymbol{R}'',\boldsymbol{R}'+\boldsymbol{R}''}=\vec{E}_{\boldsymbol{R},\boldsymbol{R}'}$, so as the DM interactions induced by them, which have a characteristic whirling pattern. The DM interactions in this case are responsible for the formation of incommensurate spin-spiral textures or skyrmion lattices~\cite{Dzyaloshinskii1964,bog1,bog2}. Such situation is realized in magnetic ferroelectrics such as PbVO$_3$~\cite{PRB2012}, BiFeO$_3$~\cite{BiFeO3}, GaV$_4$S$_8$~\cite{GaV4S8,PRB2019}, etc. 
\item[(b)] Antipolar pattern, which preserves the translational invariance (all fields are transformed to themselves by the primitive translations ${\bf t}_{1}$ and ${\bf t}_{2}$), but breaks  ${\cal I}$. The main symmetry operation in this case is fourfold rotoinversion. Such a situation is realized (though with the additional complications) in Ba$_2$CoGe$_2$O$_7$~\cite{Ba2CoG2O7.1,Ba2CoG2O7.2,Ba2CoG2O7.3,PRB2015} and Ba$_2$CuGe$_2$O$_7$~\cite{Ba2CuG2O7.1,Ba2CuG2O7.2,PRB2020}. An interesting aspect of this symmetry is the possibility of realization of antiskyrmion textures~\cite{bog3}.
\item[(c)] Antipolar zigzag pattern, satisfying the property $\vec{E}_{\boldsymbol{R}+\boldsymbol{R}'',\boldsymbol{R}'+\boldsymbol{R}''}=(-1)^{m''+n''} \vec{E}_{\boldsymbol{R},\boldsymbol{R}'}$ and doubling the unit cell as explained in Fig.~\ref{fig:E}(c). Nevertheless, these displacements respect the inversion symmetry, so as the vectors of DM interactions. As far as the DM interactions are concerned, the magnetic texture is commensurate and characterized by a canting of spins. The typical example is the weak ferromagnetism~\cite{Dzyaloshinskii_weakF}, realized in La$_2$CuO$_4$ and other materials with orthorhombically distorted perovskite lattice~\cite{La2CuO4WF,Yamaguchi}.
\end{itemize}

\par To summarize this section, in the weak FM mode, DM interactions respect the inversion symmetry. This means that the lattice is inevitably \emph{antipolar}, which results in the doubling of the unit cell. For two other modes, DM interactions are translationally invariant, but the inversion symmetry is broken. Similar behavior is expected for parameters of SO interactions in the one-orbital model. In insulators, polar and antipolar displacements lead to the realization of, respectively, ferroelectric and antiferroelectric states (so that weak ferromagnetism is a consequence of antiferroelectricity). In metals, the dipoles induced by atomic displacements are screened by conduction electrons. Nevertheless, the displacements themselves can persist and there are many examples of polar and antipolar metallic materials.

\section{\label{sec:gBloch} Generalized Bloch Theorem}
\par Suppose that two sublattices are ordered antiferromagnetically. The corresponding AFM order can be described by the propagation vector $\boldsymbol{q} = {\bf G}_{k}$, where ${\bf G}_{k} = \frac{2 \pi }{V} \varepsilon_{ijk} [{\bf T}_{i} \times {\bf T}_{j}]$ is one of the reciprocal lattice translations ($k=$ $1$, $2$, or $3$) and $\varepsilon_{ijk}$ is the antisymmetric Levi-Civita symbol, so that $\boldsymbol{q} \cdot \boldsymbol{R} = 0$ and $\pi$ (${\rm mod}$ $2\pi$) for the sublattices $1$ and $2$, respectively. 

\par Then, let us assume that the spins lie in the $xy$ plane, forming the angle $\alpha$ with the axis $x$. Corresponding N\'eel field is given by $\pm B (\cos \alpha \hat{\sigma}_{x} + \sin \alpha \hat{\sigma}_{y})$, where the signs $+$ and $-$ stand for the sublattices $1$ and $2$, respectively.

\par Consider the symmetry operations $\{ \hat{U}_{\boldsymbol{R}}|\boldsymbol{R}\}$, combining the lattice shift $\boldsymbol{R}$ with the ${\rm SU(2)}$ rotation of spins
\noindent
\begin{equation}
\hat{U}_{\boldsymbol{R}} = \frac{1}{\sqrt{2}}
\left(
\begin{array}{rr}
 1 & -1 \\
 1 &  1
\end{array}
\right)
\left(
\begin{array}{cc}
e^{\frac{i (\boldsymbol{q} \cdot \boldsymbol{R} + \alpha )}{2}} & 0 \\
0 & e^{-\frac{i( \boldsymbol{q} \cdot \boldsymbol{R} + \alpha )}{2}}
\end{array}
\right)
\label{eq:UR}
\end{equation}
\noindent such that, after the transformation, the magnetic field at both sites is pointed along $z$ and corresponding term in the Hamiltonian is given by $B \hat{\sigma}_{z}$ at both magnetic sublattices. Thus, from the viewpoint of magnetic field in the local coordinate frame, the system behaves as a ferromagnet. The result is well known in the theory of spin-spiral magnetism~\cite{Sandratskii_review,Mryasov}. Our next goal is to find whether the same applies for the transfer integrals $\hat{t}_{\boldsymbol{R},\boldsymbol{R}'} = t_{\boldsymbol{R},\boldsymbol{R}'}\hat{\mathbb{1}} + i\boldsymbol{t}_{\boldsymbol{R},\boldsymbol{R}'} \cdot \hat{\boldsymbol{\sigma}}$, where the first and second terms are, respectively, even and odd in the SO coupling. To the lowest order, $t_{\boldsymbol{R},\boldsymbol{R}'}$ is the regular transfer integral without the SO coupling, while $\boldsymbol{t}_{\boldsymbol{R},\boldsymbol{R}'}$ is induced by the SO coupling (or simply the SO coupling). Therefore, we have to find how $\hat{\boldsymbol{\sigma}}$ and $\hat{\mathbb{1}}$ are transformed by $\hat{U}_{\boldsymbol{R}}$ and $\hat{U}_{\boldsymbol{R}'}$. These transformations are given by:
\noindent
\begin{widetext}
\begin{equation}
\hat{U}_{\boldsymbol{R}}^{\phantom{\dagger}} \hat{\sigma}_{x} \hat{U}_{\boldsymbol{R}'}^{\dagger} = 
-\cos \left( \frac{\boldsymbol{q} \cdot (\boldsymbol{R}+\boldsymbol{R}')}{2} + \alpha \right) \hat{\sigma}_{z} -\sin \left( \frac{\boldsymbol{q} \cdot (\boldsymbol{R}+\boldsymbol{R}')}{2} + \alpha \right) \hat{\sigma}_{y} ,  
\end{equation}
\noindent
\begin{equation}
\hat{U}_{\boldsymbol{R}}^{\phantom{\dagger}} \hat{\sigma}_{y} \hat{U}_{\boldsymbol{R}'}^{\dagger} = 
-\sin \left( \frac{\boldsymbol{q} \cdot (\boldsymbol{R}+\boldsymbol{R}')}{2} + \alpha \right) \hat{\sigma}_{z} +\cos \left( \frac{\boldsymbol{q} \cdot (\boldsymbol{R}+\boldsymbol{R}')}{2} + \alpha \right) \hat{\sigma}_{y} , 
\end{equation}
\noindent
\begin{equation}
\hat{U}_{\boldsymbol{R}}^{\phantom{\dagger}} \hat{\sigma}_{z} \hat{U}_{\boldsymbol{R}'}^{\dagger} = 
i \sin \frac{\boldsymbol{q} \cdot (\boldsymbol{R}-\boldsymbol{R}')}{2} \hat{\mathbb{1}} +\cos \frac{\boldsymbol{q} \cdot (\boldsymbol{R}-\boldsymbol{R}')}{2}  \hat{\sigma}_{x} , 
\end{equation}
\noindent and
\noindent
\begin{equation}
\hat{U}_{\boldsymbol{R}}^{\phantom{\dagger}} \hat{\mathbb{1}} \hat{U}_{\boldsymbol{R}'}^{\dagger} =
\cos \frac{\boldsymbol{q} \cdot (\boldsymbol{R}-\boldsymbol{R}')}{2} \hat{\mathbb{1}} + i \sin \frac{\boldsymbol{q} \cdot (\boldsymbol{R}-\boldsymbol{R}')}{2} \hat{\sigma}_{x}. \label{eq:addt}
\end{equation}
\end{widetext}

\noindent Considering the transfer integrals and the SO interaction between different sublattices, we have $\frac{\boldsymbol{q} \cdot (\boldsymbol{R} \pm \boldsymbol{R}')}{2} = \frac{\pi}{2}$ (${\rm mod}$ $\pi$) and, therefore, 
\noindent
\begin{equation}
\hat{U}_{\boldsymbol{R}}^{\phantom{\dagger}} \hat{\sigma}_{x} \hat{U}_{\boldsymbol{R}'}^{\dagger} = -\sin \frac{\boldsymbol{q} \cdot (\boldsymbol{R}+\boldsymbol{R}')}{2} \left(\cos \alpha \hat{\sigma}_{y} -  \sin \alpha \hat{\sigma}_{z} \right) , \label{eq:gbtx}  
\end{equation}
\noindent
\begin{equation}
\hat{U}_{\boldsymbol{R}}^{\phantom{\dagger}} \hat{\sigma}_{y} \hat{U}_{\boldsymbol{R}'}^{\dagger} = -\sin \frac{\boldsymbol{q} \cdot (\boldsymbol{R}+\boldsymbol{R}')}{2} \left( \sin \alpha \hat{\sigma}_{y} +  \cos \alpha \hat{\sigma}_{z} \right) , \label{eq:gbty}
\end{equation}
\noindent
\begin{equation}
\hat{U}_{\boldsymbol{R}}^{\phantom{\dagger}} \hat{\sigma}_{z} \hat{U}_{\boldsymbol{R}'}^{\dagger} =
i \sin \frac{\boldsymbol{q} \cdot (\boldsymbol{R}-\boldsymbol{R}')}{2} \hat{\mathbb{1}} , \label{eq:gbtz}
\end{equation}
\noindent and
\noindent
\begin{equation}
\hat{U}_{\boldsymbol{R}}^{\phantom{\dagger}} \hat{\mathbb{1}} \hat{U}_{\boldsymbol{R}'}^{\dagger} = 
i \sin \frac{\boldsymbol{q} \cdot (\boldsymbol{R}-\boldsymbol{R}')}{2} \hat{\sigma}_{x}. \label{eq:gbt1}
\end{equation}

\par Thus, we have the following property: $\hat{U}_{\boldsymbol{R}+\boldsymbol{R}''}^{\phantom{\dagger}} \hat{\mathbb{1}} \hat{U}_{\boldsymbol{R}'+\boldsymbol{R}''}^{\dagger} = \hat{U}_{\boldsymbol{R}}^{\phantom{\dagger}} \hat{\mathbb{1}} \hat{U}_{\boldsymbol{R}'}^{\dagger}$ for any $\boldsymbol{R}''$. The same holds for $\hat{U}_{\boldsymbol{R}}^{\phantom{\dagger}} \hat{\sigma}_{z} \hat{U}_{\boldsymbol{R}'}^{\dagger}$. On the other hand, for $a=$ $x$ and $y$ we have $\hat{U}_{\boldsymbol{R}+\boldsymbol{R}''}^{\phantom{\dagger}} \hat{\sigma}_{a} \hat{U}_{\boldsymbol{R}'+\boldsymbol{R}''}^{\dagger} = (-1)^{l''+m''+n''} \, \hat{U}_{\boldsymbol{R}}^{\phantom{\dagger}} \hat{\sigma}_{a} \hat{U}_{\boldsymbol{R}'}^{\dagger}$, which is the same as the property of the SO interaction parameters $\boldsymbol{t}_{\boldsymbol{R},\boldsymbol{R}'}$ given by Eq.~(\ref{eq:tHso}). 

\par Combining these transformations of $\hat{\mathbb{1}}$ and $\hat{\boldsymbol{\sigma}}$ with $t_{\boldsymbol{R},\boldsymbol{R}'}$ and $\boldsymbol{t}_{\boldsymbol{R},\boldsymbol{R}'} = (t^{x}_{\boldsymbol{R},\boldsymbol{R}'},t^{y}_{\boldsymbol{R},\boldsymbol{R}'},t^{z}_{\boldsymbol{R},\boldsymbol{R}'})$, it is straightforward to see that 
\noindent
\begin{widetext}
\begin{equation}
t_{\boldsymbol{R}+\boldsymbol{R}'',\boldsymbol{R}'+\boldsymbol{R}''} \, \hat{U}_{\boldsymbol{R}+\boldsymbol{R}''}^{\phantom{\dagger}} \hat{\mathbb{1}} \hat{U}_{\boldsymbol{R}'+\boldsymbol{R}''}^{\dagger} = 
t_{\boldsymbol{R},\boldsymbol{R}'} \, \hat{U}_{\boldsymbol{R}}^{\phantom{\dagger}} \hat{\mathbb{1}} \hat{U}_{\boldsymbol{R}'}^{\dagger},
\end{equation}
\noindent
\begin{equation}
t^{x}_{\boldsymbol{R}+\boldsymbol{R}'',\boldsymbol{R}'+\boldsymbol{R}''} \, \hat{U}_{\boldsymbol{R}+\boldsymbol{R}''}^{\phantom{\dagger}} \hat{\sigma}_{x} \hat{U}_{\boldsymbol{R}'+\boldsymbol{R}''}^{\dagger} =
t^{x}_{\boldsymbol{R},\boldsymbol{R}'} \, \hat{U}_{\boldsymbol{R}}^{\phantom{\dagger}} \hat{\sigma}_{x} \hat{U}_{\boldsymbol{R}'}^{\dagger}, 
\end{equation}
\noindent
\begin{equation}
t^{y}_{\boldsymbol{R}+\boldsymbol{R}'',\boldsymbol{R}'+\boldsymbol{R}''} \, \hat{U}_{\boldsymbol{R}+\boldsymbol{R}''}^{\phantom{\dagger}} \hat{\sigma}_{y} \hat{U}_{\boldsymbol{R}'+\boldsymbol{R}''}^{\dagger} = 
t^{y}_{\boldsymbol{R},\boldsymbol{R}'} \, \hat{U}_{\boldsymbol{R}}^{\phantom{\dagger}} \hat{\sigma}_{y} \hat{U}_{\boldsymbol{R}'}^{\dagger}, 
\end{equation}
\noindent and
\noindent 
\begin{equation}
t^{z}_{\boldsymbol{R}+\boldsymbol{R}'',\boldsymbol{R}'+\boldsymbol{R}''} \, \hat{U}_{\boldsymbol{R}+\boldsymbol{R}''}^{\phantom{\dagger}} \hat{\sigma}_{z} 
\hat{U}_{\boldsymbol{R}'+\boldsymbol{R}''}^{\dagger} =  (-1)^{l''+m''+n''}
t^{z}_{\boldsymbol{R},\boldsymbol{R}'} \, \hat{U}_{\boldsymbol{R}}^{\phantom{\dagger}} \hat{\sigma}_{z} \hat{U}_{\boldsymbol{R}'}^{\dagger}.
\end{equation}
\end{widetext}

\par Thus, in the local coordinate frame, not only regular transfer integrals, but also two components of the SO interaction appear to be periodic on the lattice specified by the translations ${\bf t}_{1}$, ${\bf t}_{2}$, and ${\bf t}_{3}$, and transforming the sublattices $1$ and $2$ to each other. The third ($z$) component of the SO interactions is not periodic on this compact lattice and would generally require to use the two-sublattice unit cell specified by ${\bf T}_{1}$, ${\bf T}_{2}$, and ${\bf T}_{3}$. However, for the weak ferromagnets, at least one component of the SO interaction should have the same sign in all the bonds around each magnetic site and, therefore, can be eliminated via a unitary transformation on one of the sublattices~\cite{Shekhtman,arXiv2025,Kaplan}. By properly specifying the quantization axis, this component can be always chosen as $z$. The elimination procedure, which is valid to first order in the SO coupling, is explained in Appendix~\ref{sec:SOC}. In some systems, like RuO$_2$ and La$_2$CuO$_4$, having, respectively, $P4_{2}/mnm$ and $Bmab$ symmetry, the $z$ components of the SO coupling is equal to zero and, therefore, no elimination is required. At the same time, $y$ component of the SO coupling in La$_2$CuO$_4$, which is responsible for the weak ferromagnetism along $z$, should be eliminated. On the other hand, the sign-alternating components of the SO coupling do not contribute to the weak ferromagnetism and cannot be eliminated simultaneously in all the bonds~\cite{arXiv2025}.

\par For the transfer integrals operating within the sublattices, the transformation to the local coordinate frame is given by 
\noindent
\begin{equation}
\hat{U}_{\boldsymbol{R}}^{\phantom{\dagger}} \hat{\mathbb{1}} \hat{U}_{\boldsymbol{R}'}^{\dagger} =  \cos \frac{\boldsymbol{q} \cdot (\boldsymbol{R}-\boldsymbol{R}')}{2} \hat{\mathbb{1}}, \label{eq:gbt2}
\end{equation} 
\noindent which is the same for the sublattices $1$ and $2$. Therefore, these sublattices can be transformed to each other by the translations ${\bf t}_{1}$, ${\bf t}_{2}$, and ${\bf t}_{3}$ if the transfer integrals $t_{\boldsymbol{R},\boldsymbol{R}'}$ operating in the sublattice $1$ are the same as in the sublattice $2$. In this case, there will be no altermagnetic band splitting, originating from the difference of $t_{\boldsymbol{R},\boldsymbol{R}'}$ in the sublattices $1$ and $2$~\cite{arXiv2025,Roig}. However, this splitting does not seem to play a key role in AHE, which emerges already in the spin-degenerate bands~\cite{NakaOrganic,arXiv2025}. Therefore, as the first approximation, the change of the transfer integrals related to the altermagnetic band splitting can be neglected (an alternative scheme, which can be used in first-principles electronic structure calculations, will be considered in Sec.~\ref{sec:separation}). 

\par To summarize this section, the minimal model for AHE in centrosymmetric antiferromagnets can be formulated in the local coordinate frame, combining the translations ${\bf t}_{1}$, ${\bf t}_{2}$, and ${\bf t}_{3}$ with the ${\rm SU(2)}$ rotations of spins. In centrosymmetric antiferromagnets, the transformation to the local coordinate frame compensates the effect of antipolar displacements, which resulted in the doubling of the unit cell (see Sec.~\ref{sec:AFE}). The electronic Hamiltonian in this local coordinate frame becomes periodic on the lattice specified by ${\bf t}_{1}$, ${\bf t}_{2}$, and ${\bf t}_{3}$. Thus, the AFM system with the SO coupling can be effectively describe as a FM one with only one magnetic site in cell. 

\section{\label{sec:mmodel} Minimal model}
\par AHE in centrosymmetric antiferromagnets has been predicted several decades ago on phenomenological grounds~\cite{TurovBook,TurovShavrov}. In 1997, the magneto-optical effect (the ac analog of AHE) in weak ferromagnets has been studied quantitatively on the basis of first-principles electronic structure calculations~\cite{PRB1997}. Particularly, it was argued that the effect is strong and comparable with the one in regular FM state (though the net spin magnetization in the weak and regular FM states differed by two orders of magnitude). This behavior was attributed to the form of orbital magnetization, which substantially deviates from the spin one. The imaginary antisymmetric part of optical conductivity, ${\rm Im}[\sigma^{A}(\omega)]$, which was actually considered in Ref.~\cite{PRB1997}, is related to the real part, ${\rm Re}[\sigma^{A}(\omega)]$, by the Kramers-Kronig transformation, meaning that if ${\rm Im}[\sigma^{A}(\omega)]$ is finite, ${\rm Re}[\sigma^{A}(\omega)]$ is also finite. Thus, the conclusions of Ref.~\cite{PRB1997} have direct implications for the behavior of AHE in metallic regime. In this section, we introduce the minimal model, describing AHE and orbital magnetization on the microscopic level. 

\subsection{Hamiltonian}
\subsubsection{\label{sec:global} Global coordinate frame}
\par
\begin{figure}[b]
\begin{center}
\includegraphics[width=8.2cm]{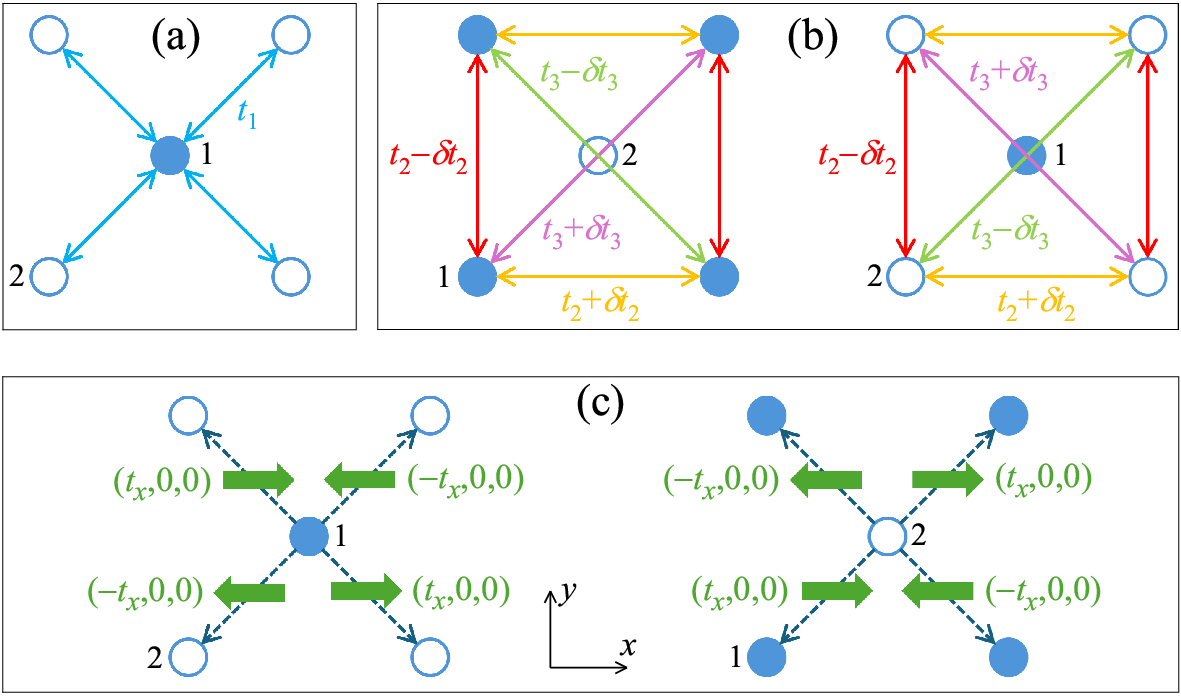} 
\end{center}
\caption{Main parameters of the model Hamiltonian for the square lattice: (a) Hoppings between first nearest neighbors; (b) Hoppings between second and third nearest neighbors; (c) Vectors of spin-orbit interaction (shown by bold arrows) around magnetic sites $1$ and $2$ after eliminating weakly ferromagnetic components. The directions of the bonds are shown by dashed arrows.}
\label{fig:model}
\end{figure}

 Main parameters of the model Hamiltonian are summarized in Fig.~\ref{fig:model}, for the square lattice~\cite{arXiv2025,Roig}. 
\noindent
\noindent They include hoppings between first ($t_{1}$), second ($t_{2}$), and third ($t_{3}$) nearest neighbors. $\delta t_{2}$ describes the orthorhombic distortion, which makes the directions $x$ and $y$ inequivalent. $\delta t_{3}$ is responsible for the altermagnetic splitting of bands. $\delta t_{2}$ and $\delta t_{3}$ operate within the sublattices. Nevertheless, $\delta t_{2}$ has the same form in the sublattices $1$ and $2$, while $\delta t_{3}$ alternates as explained in Fig.~\ref{fig:model}(b). It is assumed that SO interaction has only one sign-alternating component $x$. Two other components have the same signs in all the bonds and can be eliminated as explained in Appendix~\ref{sec:SOC}. The N\'eel field is also parallel to $x$. Then, the SO interaction and N\'eel field can be realigned along $z$, that corresponds to the global rotation of spins. The corresponding $4$$\times$$4$ spin-diagonal Hamiltonian, is given by~\cite{arXiv2025}:
\noindent
\begin{equation}
\hat{\cal H}_{\boldsymbol{k}}^{\rm g} = h^{\phantom{0}}_{\boldsymbol{k}} - \delta h^{3}_{\boldsymbol{k}} \hat{\tau}_{z} + h^{1}_{\boldsymbol{k}} \hat{\tau}_{x}  -B \hat{\tau}_{z} \hat{\sigma}_{z} + h^{\rm so}_{\boldsymbol{k}}\hat{\tau}_{y} \hat{\sigma}_{z} ,
\label{eq:HkG}
\end{equation}
\noindent where $\hat{\boldsymbol{\tau}} = (\hat{\tau}_{x},\hat{\tau}_{y},\hat{\tau}_{z})$ is a pseudospin describing interaction between two AFM sublattices~\cite{Roig}, and $h^{\phantom{0}}_{\boldsymbol{k}}$, $\delta h^{3}_{\boldsymbol{k}}$, and $h^{\rm so}_{\boldsymbol{k}}$ are the Fourier images of corresponding parameters in the  real space~\cite{arXiv2025,footnote2}. The corresponding eigenvalues are given by $\varepsilon^{\sigma}_{\boldsymbol{k},\nu} = h^{\phantom{0}}_{\boldsymbol{k}} + \nu \sqrt{(\sigma B+\delta h^{3}_{\boldsymbol{k}})^{2} + ( h^{1}_{\boldsymbol{k}})^{2} + (h^{\rm so}_{\boldsymbol{k}})^{2} }$, where $\nu = \pm$ and $\sigma = \pm$ are band and spin indices, respectively.

\par The Hamiltonian (\ref{eq:HkG}) constitutes the minimal model for altermagnets, which, with some variations, is widely used as the toy model for materials with different crystallographic symmetries~\cite{Naka,SmejkalSA,NakaOrganic,arXiv2025,Roig}. Nevertheless, below we will show that, as far as the AHE is concerned, there is even more compact model, describing this phenomenon in centrosymmetric antiferromagnets. 

\par It is instructive to consider how $\hat{\cal H}_{\boldsymbol{k}}^{\rm g}$ breaks time-reversal symmetry. The transformation rules for the different terms of Hamiltonian~(\ref{eq:HkG}) under the symmetry operations $K$, $\mathcal{T}$, $\mathcal{S}$, $\{ \mathcal{T}| {\bf t} \}$, and $\{ \mathcal{S}| {\bf t} \}$, where ${\bf t}= \left( \frac{1}{2},\frac{1}{2} \right)$ is the shift connecting two magnetic sublattices, are summarized in Table~\ref{tab:transformation}.
\noindent
\begin{table}[b]
\caption{Transformation of different terms of Hamiltonian (\ref{eq:HkG}) by the symmetry operations $K$, $\mathcal{S}$, $\mathcal{T}$, $\{ \mathcal{S}| {\bf t} \}$, and $\{ \mathcal{T}| {\bf t} \}$. The terms $\hat{\tau}_{x}$, $\hat{\tau}_{z}$, $\hat{\tau}_{y}\hat{\sigma}_{z}$, and $\hat{\tau}_{z}\hat{\sigma}_{z}$ stands for, respectively, nearest-neighbor hoppings, altermagnetic band splitting, spin-orbit interaction, and N\'eel field.}
\label{tab:transformation}  
\begin{ruledtabular}
\begin{tabular}{cccccc}
operation                    & $\hat{\tau}_{x}$ & & $\phantom{-}\hat{\tau}_{z}$              & $\phantom{-}\hat{\tau}_{y}\hat{\sigma}_{z}$                      & $\phantom{-}\hat{\tau}_{z}\hat{\sigma}_{z}$ \\
\hline
$K$                          & $\hat{\tau}_{x}$ & & $\phantom{-}\hat{\tau}_{z}$              & $-\hat{\tau}_{y}\hat{\sigma}_{z}$                                & $\phantom{-}\hat{\tau}_{z}\hat{\sigma}_{z}$ \\  
$\mathcal{S}$                & $\hat{\tau}_{x}$ & & $\phantom{-}\hat{\tau}_{z}$              & $-\hat{\tau}_{y}\hat{\sigma}_{z}$                                & $-\hat{\tau}_{z}\hat{\sigma}_{z}$           \\  
$\mathcal{T}$                & $\hat{\tau}_{x}$ & & $\phantom{-}\hat{\tau}_{z}$              & $\phantom{-}\hat{\tau}_{y}\hat{\sigma}_{z}$                      & $-\hat{\tau}_{z}\hat{\sigma}_{z}$           \\  
$\{ \mathcal{S}| {\bf t} \}$ & $\hat{\tau}_{x}$ & & $-\hat{\tau}_{z}$                        & $\phantom{-}\hat{\tau}_{y}\hat{\sigma}_{z}$                      & $\phantom{-}\hat{\tau}_{z}\hat{\sigma}_{z}$ \\  
$\{ \mathcal{T}| {\bf t} \}$ & $\hat{\tau}_{x}$ & & $-\hat{\tau}_{z}$                        & $-\hat{\tau}_{y}\hat{\sigma}_{z}$                                & $\phantom{-}\hat{\tau}_{z}\hat{\sigma}_{z}$ 
\end{tabular}
\end{ruledtabular}
\end{table}
\noindent There is no symmetry, which would transform all four terms to themselves. Thus all symmetries are broken: not only $\mathcal{T}$, but also its components $K$ and $\mathcal{S}$, both alone and in the combination with ${\bf t}$. For instance, the N\'eel field breaks $\mathcal{T}$ and $\mathcal{S}$, the SO interaction breaks $K$, $\mathcal{S}$ and $\{ \mathcal{T}| {\bf t} \}$, while the altermagnetic splitting breaks $\{ \mathcal{T}| {\bf t} \}$ and $\{ \mathcal{S}| {\bf t} \}$. Nevertheless, there are two interesting possibilities, where only one of the symmetries, either $K$ or $\{ \mathcal{S}| {\bf t} \}$, is broken, while another one is not. This means that the $\{ \mathcal{T}| {\bf t} \}$ symmetry is automatically broken, but the system still obeys certain symmetry properties governed by either $K$ or $\{ \mathcal{S}| {\bf t} \}$.
\begin{itemize}
\item[$\bullet$] The first scenario is widely discussed in the context of altermagnetism, which typically starts with the analysis of spin-splitting of bands without relativistic SO interaction~\cite{SmejkalPRX1,SmejkalPRX2,LingBai}. Then, the altermagnetic term breaks $\{ \mathcal{S}| {\bf t} \}$, so that the $\{ \mathcal{T}| {\bf t} \}$ symmetry appears to be broken even without SO interaction. Nevertheless, the Hamiltonian is real ($K$ invariant). Therefore, AHE and net orbital magnetization will vanish. Besides regular AFM order parameter, such state can be described in terms of ferroically ordered spin magnetic octupoles~\cite{McClartyRau,BhowalSpaldin,SatoHayami,OikePetersShinada}.
\item[$\bullet$] The second scenario is realized when the SO coupling is finite, but the altermagnetic splitting term identically vanishes. Then, the SO interaction, which is complex, breaks $K$. Therefore, the $\{ \mathcal{T}| {\bf t} \}$ symmetry will be also broken. Nevertheless, the Hamiltonian remains invariant under $\{ \mathcal{S}| {\bf t} \}$. Since $K$ is broken, the system will exhibit AHE and net orbital magnetization, manifesting ferro-type orbital magnetic dipolar order. 
\end{itemize}

\par In this work we explore the second possibility, which is related to AHE. In a narrow sense, the condition $\delta h^{3}_{\boldsymbol{k}}=0$ in Eq.~(\ref{eq:HkG}) defines altermagnetic nodal surfaces of spin-degenerate bands without SO coupling~\cite{SmejkalSA,Roig,ZhouRuO2}. The AHE is induced by sign-alternating component of the SO coupling, which should be parallel to the N\'eel field, as in Eq.~(\ref{eq:HkG})~\cite{NakaOrganic,arXiv2025}. However, in this configuration,  the SO interaction does not lift the spin degeneracy~\cite{Roig}. Thus, altermagnetic nodal surfaces are not essential for the AHE in centrosymmetric antiferromagnets. Moreover, there are many situations when $\delta t_{3}$ itself is small or equal to zero, either accidentally or by symmetry. This leads to the massive spin degeneracy of bands throughout the Brillouin zone even in the presence of relativistic SO coupling~\cite{NakaOrganic,arXiv2025}.

\subsubsection{\label{sec:local} Local coordinate frame}
\par Using generalized Bloch theorem, the minimal model for AHE in the reciprocal space can be formulated in the local coordinate frame via the Fourier transform, which combines the Bloch factor, $e^{i\boldsymbol{k} \cdot \boldsymbol{R}}$, associated with the translation $\boldsymbol{R}$, with additional prefactors arising from the ${\rm SU(2)}$ rotation of spins, $\hat{U}_{\boldsymbol{R}}^{\phantom{\dagger}}$, with the propagation vector $\boldsymbol{q} = {\bf G}_{k}$~\cite{Sandratskii_review,Mryasov}. Suppose that one component of the SO coupling is sign-alternating, while two other components have the same sign and can be eliminated, as explained in  Appendix~\ref{sec:SOC}. As it will become clear below, this is rather general situation, relevant to the behavior of many weak ferromagnets. The situation where there are two sign-alternating components and only one component with the same sign (or equal to zero) is also possible. It will be considered in Sec.~\ref{sec:ruo2}.

\par The $2$$\times$$2$ Hamiltonian describing two-component weak ferromagnet is given by:
\noindent
\begin{equation}
\hat{\cal H}_{\boldsymbol{k}} = h^{\phantom{0}}_{\boldsymbol{k}} \hat{\mathbb{1}}  + h^{1}_{\boldsymbol{k}} \hat{\sigma}_{x} - h^{y}_{\boldsymbol{k}}\hat{\sigma}_{y} + B \hat{\sigma}_{z},
\label{eq:Hk}
\end{equation} 
\noindent where $h^{1}_{\boldsymbol{k}}$ and $h^{y}_{\boldsymbol{k}}$ are the Fourier images of the nearest-neighbor hoppings and sign-alternating part of the SO coupling, respectively, and $h^{\phantom{0}}_{\boldsymbol{k}}$ is the Fourier image of transfer integrals within the sublattice, assuming that these transfer integrals have the same form in both AFM sublattices and the altermagnetic part, responsible for the spin splitting, can be neglected. These Fourier images can be obtained using Eqs.~(\ref{eq:gbt1}), (\ref{eq:gbty}), and (\ref{eq:gbt2}), respectively. The N\'eel field is taken in the direction of sign-alternating part of SO coupling. The directions $x$ and $y$ correspond to, respectively $\alpha = 0$ and $\pi/2$ in Eqs.~(\ref{eq:gbtx}) and (\ref{eq:gbty}). In both cases, the SO interaction is described by the matrix $\hat{\sigma}_{y}$ in the local coordinate frame, as reflected in Eq.~(\ref{eq:Hk}). According to Eq.~(\ref{eq:gbt1}), the nearest-neighbor hoppings are described by the matrix $\hat{\sigma}_{x}$. Finally, the magnetic field in the local coordinate frame is aligned along $z$ at both magnetic sites.

\par After replacing the Pauli matrices $\boldsymbol{\sigma}$ by pseudospin matrices $\boldsymbol{\tau}$, $\hat{\cal H}_{\boldsymbol{k}}$ is totally equivalent to the Hamiltonian (\ref{eq:HkG}) in the global coordinate frame for the $\sigma = -$ states and $\delta t_{3} = 0$. As will be argued in Sec.~\ref{sec:square}, the parameters $h^{\phantom{0}}_{\boldsymbol{k}}$, $h^{1}_{\boldsymbol{k}}$, and $h^{y}_{\boldsymbol{k}}\hat{\sigma}_{y}$ of these two Hamiltonian also conside. Obviously, it will yield the same anomalous Hall conductivity and net orbital magnetization. 

\par The Hamiltonian has two eigenvalues:
\noindent
\begin{equation}
\varepsilon^{\pm}_{\boldsymbol{k}} = h^{\phantom{0}}_{\boldsymbol{k}} \mp \sqrt{B^{2} + ( h^{1}_{\boldsymbol{k}})^{2} + (h^{y}_{\boldsymbol{k}})^{2} } \equiv h^{\phantom{0}}_{\boldsymbol{k}} \mp A_{\boldsymbol{k}},
\label{eq:ekpm}
\end{equation}
\noindent for the majority ($\varepsilon^{+}_{\boldsymbol{k}}$) and minority ($\varepsilon^{-}_{\boldsymbol{k}}$) spin states in the local coordinate frame. Searching corresponding to them eigenvectors in the form 
\noindent
\begin{equation}
| u^{+}_{\boldsymbol{k}} \rangle = \left(
\begin{array}{c}
\cos \theta_{\boldsymbol{k}} e^{i \phi_{\boldsymbol{k}}} \\
\sin \theta_{\boldsymbol{k}}
\end{array}
\right) \equiv | u^{\phantom{+}}_{\boldsymbol{k}} \rangle
\label{eq:up}
\end{equation}
\noindent and
\noindent
\begin{equation}
| u^{-}_{\boldsymbol{k}} \rangle = \left(
\begin{array}{c}
-\sin \theta_{\boldsymbol{k}} \\
\cos \theta_{\boldsymbol{k}} e^{-i \phi_{\boldsymbol{k}}}
\end{array}
\right) ,
\end{equation}
\noindent it is straightforward to find that
\noindent
\begin{equation}
- A_{\boldsymbol{k}} \cos 2 \theta_{\boldsymbol{k}} = B 
\end{equation}
\noindent and
\begin{equation}
-A_{\boldsymbol{k}} \sin 2 \theta_{\boldsymbol{k}} e^{i \phi_{\boldsymbol{k}}} = h^{1}_{\boldsymbol{k}} + i h^{y}_{\boldsymbol{k}},
\end{equation}
\noindent where
\noindent
\begin{equation}
\theta_{\boldsymbol{k}}  = \frac{1}{2} \arctan \frac{ \sqrt{( h^{1}_{\boldsymbol{k}})^{2} + (h^{y}_{\boldsymbol{k}})^{2}}} {B} 
\end{equation}
\noindent ($0 \le \theta_{\boldsymbol{k}} < \pi$) and
\noindent
\begin{equation}
\phi_{\boldsymbol{k}} = \arctan \left( \frac{h^{y}_{\boldsymbol{k}}}{h^{1}_{\boldsymbol{k}}} \right) 
\label{eq:phik}
\end{equation}
\noindent ($0 \le \phi_{\boldsymbol{k}} < 2\pi$).

\subsection{Anomalous Hall Conductivity}
\par Using the above expressions and considering the situation when $B$ is sufficiently large, so that the majority-spin band is partly occupied, while the minority-spin band is empty, it is straightforward to obtain that the Berry curvature, $\Omega^{c}(\boldsymbol{k}) = -2 {\rm Im} \left\langle \partial_{k_{a}} u_{\boldsymbol{k}} | \partial_{k_{b}} u_{\boldsymbol{k}} \right\rangle$~\cite{NagaosaRevModPhys}, becomes
\noindent
\begin{equation}
\Omega_{\boldsymbol{k}}^{c} = \sin 2 \theta_{\boldsymbol{k}} \left( \partial_{k_{a}} \theta_{\boldsymbol{k}} \partial_{k_{b}} \phi_{\boldsymbol{k}} - \partial_{k_{a}} \phi_{\boldsymbol{k}} \partial_{k_{b}} \theta_{\boldsymbol{k}} \right) ,
\label{eq:Berryc0}
\end{equation}
\noindent where
\noindent
\begin{equation}
\sin 2 \theta_{k} = -\frac{\sqrt{(h^{1}_{\boldsymbol{k}})^{2} + (h^{y}_{\boldsymbol{k}})^{2}}}{A_{\boldsymbol{k}}},
\label{eq:sin2t}
\end{equation}
\noindent
\begin{equation}
\partial_{k_{a}} \theta_{\boldsymbol{k}} = \frac{1}{2} \frac{(h^{1}_{\boldsymbol{k}} \partial_{k_{a}} h^{1}_{\boldsymbol{k}} + h^{y}_{\boldsymbol{k}} \partial_{k_{a}} h^{y}_{\boldsymbol{k}})B }{A_{\boldsymbol{k}}^{2} \sqrt{(h^{1}_{\boldsymbol{k}})^{2} + (h^{y}_{\boldsymbol{k}})^{2}}},
\end{equation}
\noindent 
\begin{equation}
\partial_{k_{a}} \phi_{\boldsymbol{k}} = \frac{h^{1}_{\boldsymbol{k}} \partial_{k_{a}} h^{y}_{\boldsymbol{k}} - h^{y}_{\boldsymbol{k}} \partial_{k_{a}} h^{1}_{\boldsymbol{k}}}{(h^{1}_{\boldsymbol{k}})^{2} + (h^{y}_{\boldsymbol{k}})^{2}} ,
\label{eq:daphi}
\end{equation}
\noindent and $abc$ is an even permutation of $xyz$.

\par Then, after some algebra, $\Omega_{\boldsymbol{k}}^{c}$ can be further rearranged as
\noindent
\begin{equation}
\Omega_{\boldsymbol{k}}^{c} = \frac{B}{2 A^{3}_{\boldsymbol{k}}} \left( \partial_{k_{a}} h^{y}_{\boldsymbol{k}} \partial_{k_{b}} h^{1}_{\boldsymbol{k}} - \partial_{k_{a}} h^{1}_{\boldsymbol{k}} \partial_{k_{b}} h^{y}_{\boldsymbol{k}} \right).
\label{eq:Berryc}
\end{equation} 

\par The anomalous Hall conductivity is given by the Brillouin zone (${\rm BZ}$) integral~\cite{NagaosaRevModPhys}
\begin{displaymath}
\sigma_{ab} =  - \frac{1}{V} \int_{\rm BZ} \frac{d \boldsymbol{k}}{\Omega_{\rm BZ}} f_{\boldsymbol{k}} \Omega^{c}_{\boldsymbol{k}},  
\end{displaymath}
\noindent where $\Omega_{\rm BZ} = \frac{(2 \pi)^{3}}{V}$ is the ${\rm BZ}$ volume and $f_{\boldsymbol{k}}$ is the Fermi-Dirac distribution function for $\varepsilon^{+}_{\boldsymbol{k}}$. In the large $B$ limit, $\Omega^{c}_{\boldsymbol{k}} \sim \frac{1}{B|B|}$~\cite{arXiv2025}. Therefore, the Taylor expansion of $\sigma_{ab}$ in terms of $B$ is not meaningful and only the expansion in the directions of the N\'eel field is generally well defined~\cite{LiuPRX}. However, it does not contradict to phenomenological theories~\cite{Dzyaloshinskii_weakF,DzyaloshinskiiPM,DzyaloshinskiiME,TurovBook,TurovShavrov,McClartyRau,TurovUFN}, where absolute value of the AFM order parameter (and other quantities) is generally unknown and these theories establish only the directional dependence.

\par The magneto-optical effect (the ac analog of AHE) can be also evaluated in the local coordinate frame by considering the optical transitions between the $u^{-}_{\boldsymbol{k}}$ and $u^{+}_{\boldsymbol{k}}$ states.

\subsection{Orbital Magnetization}
\par The orbital magnetization along $c$ is given by the ${\rm BZ}$ integral of
\noindent
\begin{equation}
{\cal M}^{c}_{\boldsymbol{k}} =  {\rm Im} \left\langle \partial_{k_{a}} u_{\boldsymbol{k}} | \hat{\cal H}_{\boldsymbol{k}} + \varepsilon^{+}_{\boldsymbol{k}} -2\varepsilon_{\rm F}  | \partial_{k_{b}} u_{\boldsymbol{k}} \right\rangle 
\end{equation}
\noindent with the Fermi-Dirac distribution function~\cite{Thonhauser,Shi}. Using the explicit form of $\hat{\cal H}_{\boldsymbol{k}}$, $\varepsilon^{+}_{\boldsymbol{k}}$, and $| u_{\boldsymbol{k}} \rangle$, ${\cal M}^{c}_{\boldsymbol{k}}$ can be rearranged as the sum of four contributions~\cite{arXiv2025}:
\noindent
\begin{widetext}
\begin{equation}
{\cal M}_{\boldsymbol{k},{\rm I}}^{c} = -\frac{1}{2}\sin 2 \theta_{\boldsymbol{k}} \left(2h^{\phantom{0}}_{\boldsymbol{k}} - A_{\boldsymbol{k}} \right) \left( \partial_{k_{a}} \theta_{\boldsymbol{k}} \partial_{k_{b}} \phi_{\boldsymbol{k}} - \partial_{k_{a}} \phi_{\boldsymbol{k}} \partial_{k_{b}} \theta_{\boldsymbol{k}} \right)  ,
\end{equation}
\noindent
\begin{equation}
{\cal M}_{\boldsymbol{k},{\rm II}}^{c} =  \left( \varepsilon_{\rm F} - \frac{B}{2} \right)  {\Omega}_{\boldsymbol{k}}^{c} ,
\end{equation}
\noindent
\begin{eqnarray}
{\cal M}_{\boldsymbol{k},{\rm III}}^{c} = h^{1}_{\boldsymbol{k}} \cos^{2} \theta_{\boldsymbol{k}} \cos \phi_{\boldsymbol{k}} \left( \partial_{k_{a}}\theta_{\boldsymbol{k}} \partial_{k_{b}}\phi_{\boldsymbol{k}} - \partial_{k_{a}}\phi_{\boldsymbol{k}} \partial_{k_{b}}\theta_{\boldsymbol{k}}   \right) ,
\end{eqnarray}
\noindent and 
\noindent
\begin{eqnarray}
{\cal M}_{\boldsymbol{k},{\rm IV}}^{c} = h^{y}_{\boldsymbol{k}} \cos^{2} \theta_{\boldsymbol{k}} \sin \phi_{\boldsymbol{k}} \left( \partial_{k_{a}}\theta_{\boldsymbol{k}} \partial_{k_{b}}\phi_{\boldsymbol{k}} - \partial_{k_{a}}\phi_{\boldsymbol{k}} \partial_{k_{b}}\theta_{\boldsymbol{k}}   \right)  .
\end{eqnarray} 
\end{widetext}

\par Then, using Eq.~(\ref{eq:Berryc0}), all these terms can be related to $\Omega_{\boldsymbol{k}}^{c}$ as
\noindent
\begin{equation}
{\cal M}_{\boldsymbol{k},{\rm I}}^{c} = -\frac{1}{2} \left(2h^{\phantom{0}}_{\boldsymbol{k}} - A_{\boldsymbol{k}} \right) \Omega_{\boldsymbol{k}}^{c} ,
\end{equation}
\noindent
\begin{equation}
{\cal M}_{\boldsymbol{k},{\rm III}}^{c} =  \frac{h^{1}_{\boldsymbol{k}} \cos \phi_{\boldsymbol{k}}\cos^{2} \theta_{\boldsymbol{k}}}{\sin 2 \theta_{\boldsymbol{k}}} \Omega_{\boldsymbol{k}}^{c} ,
\end{equation}
\noindent and
\noindent
\begin{equation}
{\cal M}_{\boldsymbol{k},{\rm IV}}^{c} =  \frac{h^{y}_{\boldsymbol{k}} \sin \phi_{\boldsymbol{k}}\cos^{2} \theta_{\boldsymbol{k}}}{\sin 2 \theta_{\boldsymbol{k}}} \Omega_{\boldsymbol{k}}^{c} .
\end{equation}

\par Noting that
\noindent
\begin{equation} 
\frac{\cos^{2} \theta_{\boldsymbol{k}}}{\sin 2 \theta_{\boldsymbol{k}}} = \frac{1}{2} \frac{B-A_{\boldsymbol{k}}}{\sqrt{(h^{1}_{\boldsymbol{k}})^{2} + (h^{y}_{\boldsymbol{k}})^{2}}}
\end{equation}
\noindent and $h^{1}_{\boldsymbol{k}} \cos \phi_{\boldsymbol{k}} + h^{y}_{\boldsymbol{k}} \sin \phi_{\boldsymbol{k}} = \sqrt{(h^{1}_{\boldsymbol{k}})^{2} + (h^{y}_{\boldsymbol{k}})^{2}}$, it is straightforward to find that ${\cal M}_{\boldsymbol{k},{\rm III}}^{c} + {\cal M}_{\boldsymbol{k},{\rm IV}}^{c} = \frac{1}{2}\left( B-A_{\boldsymbol{k}} \right) \Omega_{\boldsymbol{k}}^{c}$. Summing up all four terms, one can get the final expression:
\noindent
\begin{equation}
{\cal M}^{c}_{\boldsymbol{k}} =  \left( \varepsilon_{\rm F} - h^{\phantom{0}}_{\boldsymbol{k}} \right) \Omega_{\boldsymbol{k}}^{c}. 
\label{eq:OMk}
\end{equation}

\par Thus, the orbital magnetization is 
\begin{displaymath}
{\cal M}^{c} =  \int_{\rm BZ} \frac{d \boldsymbol{k}}{\Omega_{\rm BZ}} f_{\boldsymbol{k}} \left( \varepsilon_{\rm F} - h^{\phantom{0}}_{\boldsymbol{k}} \right) \Omega_{\boldsymbol{k}}^{c} .
\end{displaymath}

\section{\label{sec:examples} Examples}

\subsection{\label{sec:square} Square perovskite lattice}
\par As the first example, let us consider the single layer of orthorhombically distorted perovskites with the $Pbnm$ or $Bmab$ symmetry~\cite{arXiv2025}. The lattice is specified by the primitive translations ${\bf T}_{1} = (1,0,0)$ and ${\bf T}_{2} = (0,1,0)$, and there are two sublattices, which are connected by the vectors ${\bf t}_{1} = (\frac{1}{2},\frac{1}{2}) \equiv {\bf t}$ and ${\bf t}_{2} = (-\frac{1}{2},\frac{1}{2})$. The symmetry operation transforming two sublattices to each other is $\{ {\cal C}_{2x} | {\bf t} \}$, meaning that $x$ components of the SO coupling is sign-alternating. For the space group $Bmab$, the lattice is additionally invariant under the mirror reflection $y \to -y$. 

\par The minimal model was introduced in Sec.~\ref{sec:global} and explained in Fig.~\ref{fig:model}. We would like to emphasize again that the SO interaction obeys the following property, being the consequence of the inversional invariance: $\hat{\cal H}^{\rm so}_{\boldsymbol{R}+\boldsymbol{R}'',\boldsymbol{R}'+\boldsymbol{R}''} = (-1)^{m''+n''} \, \hat{\cal H}^{\rm so}_{\boldsymbol{R},\boldsymbol{R}'}$. In the global frame the SO interaction is not periodic on the lattice, specified by ${\bf t}_{1}$ and ${\bf t}_{2}$. However, it can be made periodic in the local frame as explained in Secs.~\ref{sec:basic}-\ref{sec:gBloch}. We do not consider the altermagnetic deformation of the third-neighbor hoppings, which is identically equal to zero for the $Bmab$ symmetry of La$_2$CuO$_4$ and expected to be small for other materials~\cite{arXiv2025}. For La$_2$CuO$_4$, this is the consequence of the mirror reflection $y \to -y$, which makes the bonds with $t_{3} \pm \delta t_{3}$ in Fig.~\ref{fig:model}(b) equivalent. For the one-orbital model, which is widely used in the theory of superconductivity~\cite{ImadaPRB,ImadaPRX}, it leads to $\delta t_{3} = 0$. In a more general multiorbital model, $\delta t_{3}$ can be finite, provided that it operate between \emph{different} orbitals belonging to odd and even representations of the point group. Nevertheless, the band splitting, which is driven by this $\delta t_{3}$, is small~\cite{SmejkalPRX1}. 

\par Thus, one can readily find the Hamiltonian (\ref{eq:Hk}) in the local coordinate frame, where $h^{\phantom{0}}_{\boldsymbol{k}}$, $h^{1}_{\boldsymbol{k}}$, and $h^{y}_{\boldsymbol{k}}$ are obtained by combining the Fourier transforms with the matrix elements of $\hat{\mathbb{1}}$ and $\hat{\sigma}_{x}$ given by Eqs.~(\ref{eq:gbt1}) and (\ref{eq:gbtx}), respectively. Moreover, it is convenient to shift the $\boldsymbol{k}$-mesh: $(k_{x},k_{y}) \to (k_{x} + \pi,k_{y})$. Altogether, this yields: $h^{\phantom{0}}_{\boldsymbol{k}} = 2t_{2} (\cos k_{x} + \cos k_{y}) + 2 \delta t_{2}(\cos k_{x} - \cos k_{y}) + 4t_{3} \cos k_{x} \cos k_{y}$, $h^{1}_{\boldsymbol{k}} = 4 t_{1} \cos \frac{k_{x}}{2} \cos \frac{k_{y}}{2}$, and $h^{y}_{\boldsymbol{k}} = -4t_{x} \sin \frac{k_{x}}{2} \sin \frac{k_{y}}{2}$, which coincide with the parameters reported in Ref.~\cite{arXiv2025} in the global coordinate frame~\cite{footnote2}. 

\par Berry curvature can be written in the compact form, using Eq.~(\ref{eq:Berryc}):
\noindent
\begin{equation}
\Omega_{\boldsymbol{k}}^{z} = \frac{Bt_{1}t_{x}}{A_{\boldsymbol{k}}^{3}} \left( \cos k_{x} - \cos k_{y} \right). 
\end{equation} 

\par
\begin{figure}[t]
\begin{center}
\includegraphics[width=8.2cm]{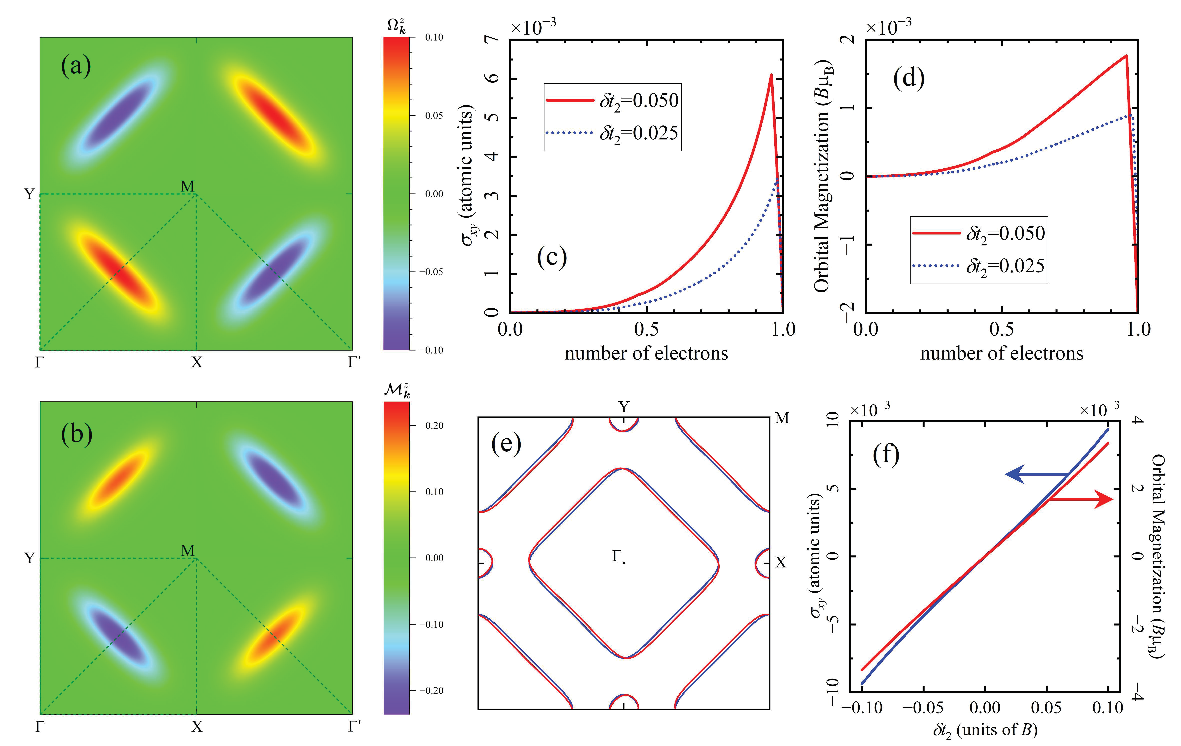}
\end{center}
\caption{Results for the square lattice model with the parameters (unless it is specified otherwise) $t_{1}=-1$, $t_{2}= -t_{3} =0.1$, and $\delta t_{2}= -t_{x} = 0.05$ (all are in units of $B$): (a) Berry curvature, $\Omega^{z}_{\boldsymbol{k}}$, in the first Brillouin zone; (b) Similar plot for orbital magnetization, ${\cal M}^{z}_{\boldsymbol{k}}$, corresponding to $n_{\rm el}=0.5$ electrons; Band filling dependence of (c) the anomalous Hall conductivity, $\sigma_{xy}$, and (d) the orbital magnetization, ${\cal M}^{z}$; (e) Fermi surface at $n_{\rm el}=0.5$ with (red) and without (blue) the orthorhombic strain $\delta t_{2}$; and (f) The orthorhombic strain dependence of $\sigma_{xy}$ and ${\cal M}^{z}$ for $n_{\rm el}=0.9$.} 
\label{fig:square}
\end{figure}

 The results are summarized in Fig.~\ref{fig:square}.
\noindent
\noindent In comparison with calculations in the global coordinate frame~\cite{arXiv2025}, the generalized Bloch theorem allows us to use smaller unit cell. Therefore, the first Brillouin zone becomes larger, where $\Omega^{z}_{\boldsymbol{k}}$ has two positive and two negative areas [Fig.~\ref{fig:square}(a)], which tend to compensate each other. Therefore, in order to obtain finite $\sigma_{xy}$ and ${\cal M}^{z}$, it is essential to consider the orthorhombic strain, $\delta t_{2}$~\cite{NakaOrganic,arXiv2025}. Although the Berry curvature does not depend on $\delta t_{2}$, the latter deforms the Fermi surface, as shown in Fig.~\ref{fig:square}(e), so that the positive and negative areas of $\Omega^{z}_{\boldsymbol{k}}$ will contribute to $\sigma_{xy}$ with different weights, resulting in finite $\sigma_{xy}$. According to Eq.~(\ref{eq:OMk}), in the case of ${\cal M}^{z}$, there is an additional modulation of $\Omega^{z}_{\boldsymbol{k}}$ by $h^{\phantom{0}}_{\boldsymbol{k}}$, which makes inequivalent the positive and negative areas of ${\cal M}^{z}_{\boldsymbol{k}}$ [see Fig.~\ref{fig:square}(b)]. Nevertheless the effect is small if $\delta t_{2}$ is small. 

\par Finally, AHE and orbital magnetization can be efficiently controlled by the orthorhombic strain, where the transition from tensile ($\delta t_{2}<0$) to compressive ($\delta t_{2}>0$) strain changes the sign of $\sigma_{xy}$ and ${\cal M}^{z}$ [see Fig.~\ref{fig:square}(f)]. This effect has a clear similarity with the piezomagnetizm~\cite{JunweiLiuNC,JunweiLiuPRX,arXiv2025}. For small $\delta t_{2}$, the contribution $\varepsilon_{\rm F} \Omega^{z}_{\boldsymbol{k}}$ to the orbital magnetization dominates and ${\cal M}^{z} \sim \sigma_{xy}$.

\subsection{\label{sec:CuF2} VF$_4$ and CuF$_2$}
\par As the next example, we consider the minimal model for VF$_4$ and CuF$_2$, which are regarded as altermagnetic candidates~\cite{LingBai}. Both materials crystallize in the monoclinic structure (the space group $P2_{1}/c$, No. 14) with two formula units in the primitive cell~\cite{VF4,CuF2}. V$^{4+}$ has one $d$ electron, while Cu$^{2+}$ has one $d$ hole. According to the local-density approximation (LDA), the electronic structure of VF$_4$ and CuF$_2$ near the Fermi level is featured by well isolated half-filled band. Therefore, in the first approximation, the properties of these materials can be described by the one-orbital model with the parameters derived from the first-principles calculations. This is basically an extension of the previous considerations to the case of lower crystallographic symmetry. The details of electronic structure calculations and construction of the model are described in Appendix~\ref{sec:CuF2model}. The main results are summarized in Table~\ref{tab:CuF2}.
\noindent
\begin{table}[b]
\caption{Experimental lattice parameters ($a$, $b$, and $c$ are in~\AA, $\beta$ is in degrees) and main parameters of the one-orbital model for VF$_4$ and CuF$_2$: the nearest-neighbor hopping $t_{1}$ (in meV), the sign-alternating component of the spin-orbit coupling $t_{y}$ (in meV), and on-site Coulomb repulsion $U$ (in eV). The lattice parameters are taken from Refs.~\cite{VF4} and \cite{CuF2}.}
\label{tab:CuF2}
\begin{ruledtabular}
\begin{tabular}{cccccccc} 
        & $a$       & $b$       & $c$       & $\beta$        & $t_{1}$       & $t_{y}$       & $U$     \\
\hline
VF$_4$  & $5.381$   & $5.170$   & $5.340$   &  $59.74$       & $-100.66$     & $0.71$        &  $3.2$  \\
CuF$_2$ & $3.297$   & $4.562$   & $4.616$   &  $83.29$       & $-173.73$     & $5.02$        &  $4.0$  
\end{tabular}
\end{ruledtabular}
\end{table}

\par The primitive translations are ${\bf T}_{1} = (a \sin \beta, 0, a \cos \beta)$, ${\bf T}_{2} = (0, b, 0)$, and ${\bf T}_{3} = (0, 0, c)$, both for VF$_4$ and CuF$_2$. Two sublattices are transformed to each other by the symmetry operation $\{ {\cal C}_{2y} | {\bf t} \}$. Therefore, the $x$ and $z$ components of SO interaction between the nearest neighbor have the same sign in all the bonds and can be eliminated (see Appendix~\ref{sec:SOC}). The $y$ component is sign-alternating. The N\'eel field parallel to the $y$ axis will result in the weak ferromagnetism along $x$ and $z$.

\par The atoms in the nearest neighborhood of each atomic site are located at ${\bf t}_{3}$, $-$${\bf t}_{3}$, ${\bf t}_{3}$$-$${\bf T}_{2}$, and $-$${\bf t}_{3}$$+$${\bf T}_{2}$. The AFM propagation vector can be taken as $\boldsymbol{q} = \boldsymbol{G}_{2}$, so that $\sin \frac{\boldsymbol{q} \cdot {\bf t}_{3}}{2} = 1$ and $\sin \frac{\boldsymbol{q} \cdot ({\bf t}_{3}-{\bf T}_{2})}{2} = - 1$. Furthermore, in the first two bonds, the SO coupling is $it_{y}$, while in the second two bonds it is $-it_{y}$. Then, using Eqs.~(\ref{eq:gbt1}) and (\ref{eq:gbty}) for $\alpha = \pi / 2$, and shifting the $\boldsymbol{k}$-mesh $\boldsymbol{k} \to \boldsymbol{k} + \frac{\boldsymbol{q}}{2}$, one can find $h^{1}_{\boldsymbol{k}} = 4 t_{1} \cos \frac{\boldsymbol{k} \cdot ({\bf T}_{1}-{\bf T}_{3})}{2} \cos \frac{\boldsymbol{k} \cdot {\bf T}_{2}}{2}$ and $h^{y}_{\boldsymbol{k}} = 4 t_{y} \sin \frac{\boldsymbol{k} \cdot ({\bf T}_{1}-{\bf T}_{3})}{2} \sin \frac{\boldsymbol{k} \cdot {\bf T}_{2}}{2}$. Finally, $h^{\phantom{0}}_{\boldsymbol{k}}$ in Eq.~(\ref{eq:Hk}) is the Fourier image of transfer integrals between atoms of one magnetic sublattice and assuming that the they are the same in both sublattices (otherwise the generalized Bloch theorem does not apply). Technically, $h^{\phantom{0}}_{\boldsymbol{k}}$ is obtained by averaging the transfer integrals over two magnetic sublattices. The details and applicability of this approximation are discussed in Appendix~\ref{sec:CuF2model}. Briefly, the hopping parameters $\delta t_{\boldsymbol{R},\boldsymbol{R}'}$ responsible for the altermagnetic spin-splitting of bands are small in VF$_4$, but can be comparable with regular hoppings $t_{\boldsymbol{R},\boldsymbol{R}'}$ in the case of CuF$_2$. In the strong coupling limit, the contribution of degenerate bands to AHE is of the order of $\frac{t_{1}t_{y}}{B|B|}$, where $B \approx \frac{U}{2}$. The spin splitting yields the additional contribution $\sim \frac{2t_{1}t_{y} \delta t_{\boldsymbol{R},\boldsymbol{R}'}}{|B|^{3}}$, which is smaller by the factor $\frac{2 \delta t_{\boldsymbol{R},\boldsymbol{R}'}}{|B|}$ ~\cite{arXiv2025}. The realistic parameters $|\delta t_{\boldsymbol{R},\boldsymbol{R}'}| < 30$ meV (Appendix~\ref{sec:CuF2model}) and $U = 4$ eV (Table~\ref{tab:CuF2}) yield the following estimate $\frac{2 | \delta t_{\boldsymbol{R},\boldsymbol{R}'}|}{|B|} < 0.03$. Therefore, the altermagnetic contribution to AHE, originating from the spin-splitting of bands, is expected to be small. 

\par Then, using Eq.~(\ref{eq:Berryc}), it is straightforward to find that
\noindent
\begin{widetext}
\begin{equation}
\Omega_{\boldsymbol{k}}^{x} = \frac{Bt_{1}t_{y}}{A_{\boldsymbol{k}}^{3}} (a\cos \beta - c)b \left\{ \cos \boldsymbol{k} \cdot ({\bf T}_{1}-{\bf T}_{3}) - \cos \boldsymbol{k} \cdot {\bf T}_{2} \right\} ,
\end{equation}
\noindent $\Omega_{\boldsymbol{k}}^{y} = 0$, and
\noindent
\begin{equation}
\Omega_{\boldsymbol{k}}^{z} = -\frac{Bt_{1}t_{y}}{A_{\boldsymbol{k}}^{3}} ab\sin \beta \left\{ \cos \boldsymbol{k} \cdot ({\bf T}_{1}-{\bf T}_{3}) - \cos \boldsymbol{k} \cdot {\bf T}_{2} \right\} .
\end{equation}
\noindent Therefore, the ratio 
\noindent
\begin{equation}
\frac{\Omega_{\boldsymbol{k}}^{x}}{\Omega_{\boldsymbol{k}}^{z}} = \frac{c-a\cos \beta}{a \sin \beta}
\label{eq:CuF2scaling}
\end{equation}
\end{widetext}
\noindent is controlled by the geometrical factor, depending only on the lattice parameters. 

\par
\begin{figure}[b]
\begin{center}
\includegraphics[width=8.2cm]{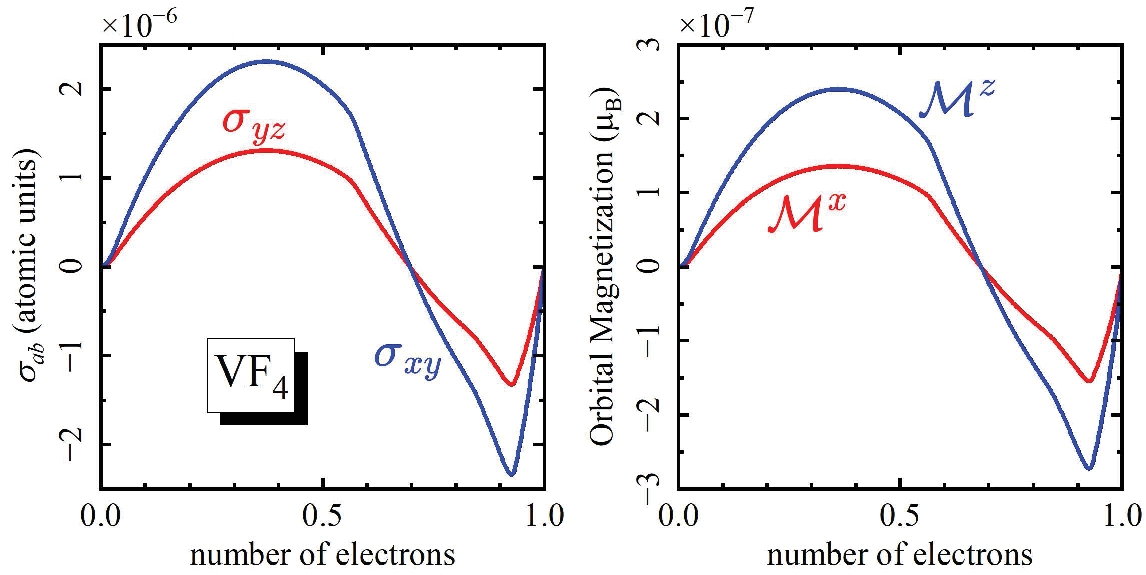} 
\end{center}
\begin{center}
\includegraphics[width=8.2cm]{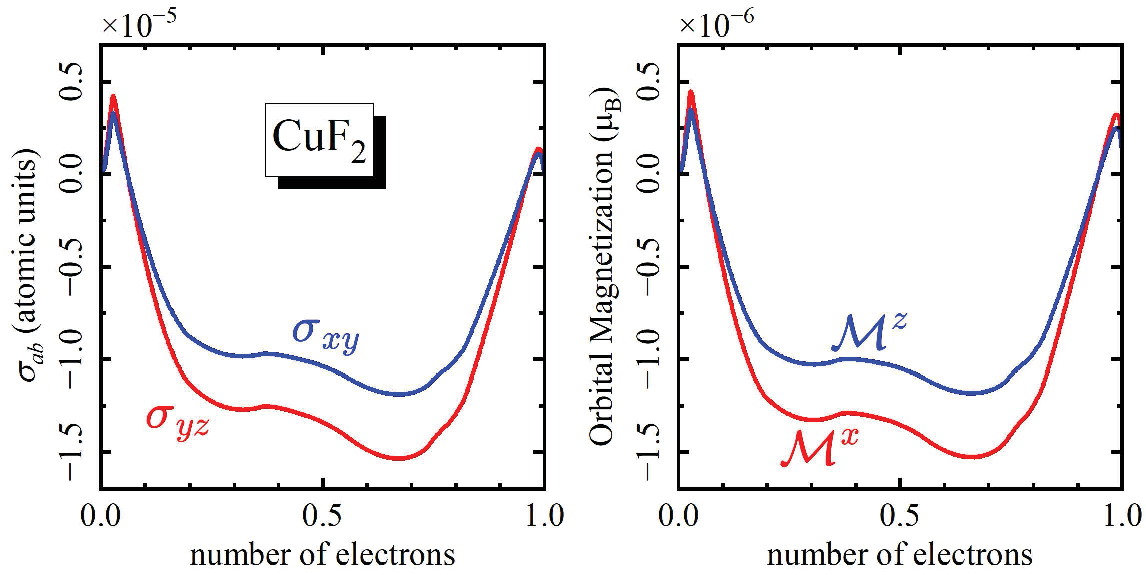} 
\end{center}
\caption{Band filling dependence of anomalous Hall conductivity and orbital magnetization in VF$_4$ and CuF$_2$.} 
\label{fig:CuF2AHE}
\end{figure}

 These tendencies are clearly seen in the behavior of AHE and orbital magnetization (Fig.~\ref{fig:CuF2AHE}).
\noindent
\noindent The nonvanishing $yz$ and $xy$ components of the Hall conductivity are related to each other by a scaling transformation, which is expected from Eq.~(\ref{eq:CuF2scaling}). The same holds for the $x$ and $z$ components of the orbital magnetization. Furthermore, the shape of orbital magnetization basically repeats the one of AHE, as expected from Eq.~(\ref{eq:OMk}).

\par Finally, we would like to comment on the possibility of practical realization of AHE (under doping) or magneto-optical effect in VF$_4$ and CuF$_2$. One important question is whether $\boldsymbol{q} = \boldsymbol{G}_{2} \equiv (0,1,0)$ corresponds to the true magnetic ground states of VF$_4$ and CuF$_2$. For instance, from the viewpoint of symmetry, a similar material CuO could be classified as a potential altermagnet: the space group is $C2/c$, where there are two Cu sublattices. Each sublattice is transformed to itself by $\mathcal{I}$, while different sublattices are transformed to each other by $\{\mathcal{C}_{2y} | \bf{t} \}$. However, the magnetic ground state of CuO corresponds to $\boldsymbol{q} \approx (\frac{1}{2},0,\frac{1}{2})$, meaning that it is a noncollinear multiferroic~\cite{Kimura_CuO}. This tendency is successfully explained by the behavior of superexchange interactions~\cite{PRL2022}. The same situation is expected in VF$_4$ and CuF$_2$, where the analysis of superexchange interactions also predicts the multiferroic $\boldsymbol{q} \approx (\frac{1}{2},0,\frac{1}{2})$ ground state. 

\subsection{\label{sec:ruo2} ${\rm RuO_{2}}$-type systems}
\par
\begin{figure}[t]
\begin{center}
\includegraphics[width=8.2cm]{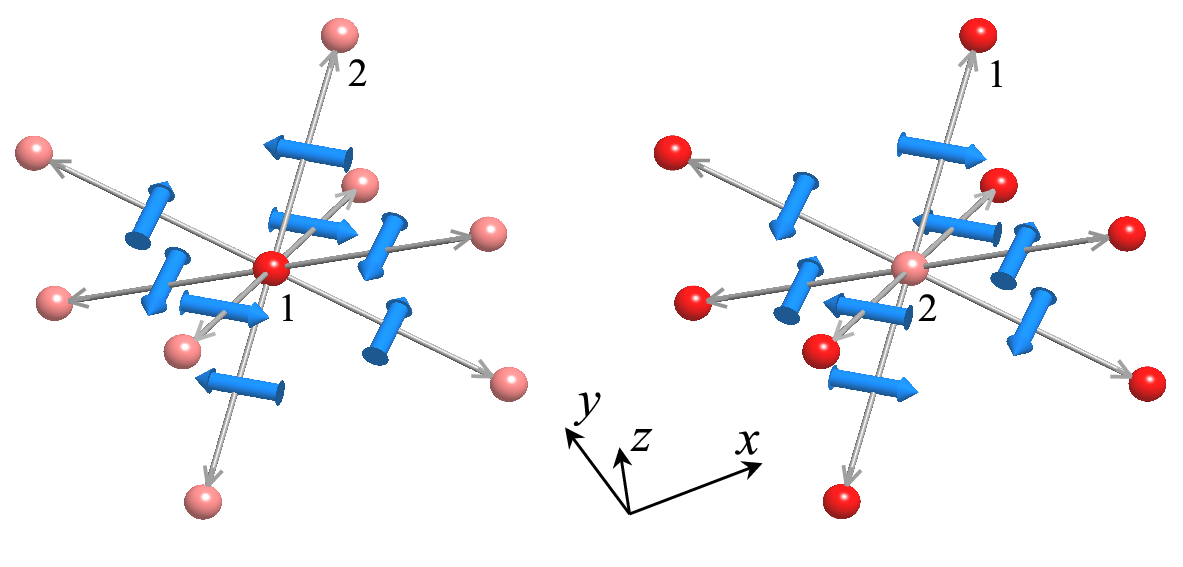} 
\end{center}
\caption{Spin-orbit interactions $\boldsymbol{t}_{\boldsymbol{R},\boldsymbol{R}'}$  (denoted by bold blue arrows) around two magnetic sites in the $P4_{2}/mnm$ structure. The vectors $\boldsymbol{t}_{\boldsymbol{R},\boldsymbol{R}'}$ replicate the form of Dzyaloshinskii-Moriya interactions. The corresponding bond directions are shown by grey arrows.} 
\label{fig:RuO2}
\end{figure}

 RuO$_2$ is viewed as a canonical altermagnetic material~\cite{SmejkalSA}. It crystallizes in the rutile-type body-centered tetragonal (bct) structure with two sublattices (the space group  $P4_{2}/mnm$, No. 136). The atomic positions in both sublattices can be formally generated by the vectors ${\bf t}_{1} = (-\frac{1}{2},\frac{1}{2},\frac{1}{2})$, ${\bf t}_{2} = (\frac{1}{2},-\frac{1}{2},\frac{1}{2})$, and ${\bf t}_{3} = (\frac{1}{2},\frac{1}{2},-\frac{1}{2})$, in units of $a$ and $c$ ($c/a<1$ being the tetragonal distortion along $z$). The additional complication arising in this case is that there are two sign-alternating components of the SO coupling in the nearest bonds: $x$ and $y$, as explained in Fig.~\ref{fig:RuO2}. 
\noindent
\noindent The third ($z$) component is identically equal to zero. Thus, formally speaking, there is no weak spin ferromagnetism arising from DM interactions, which is believed to be a unique aspect of RuO$_2$ and other AFM materials with the $P4_{2}/mnm$ symmetry. However, in other materials, the weakly FM component can be eliminated in the local coordinate frame (see Appendix~\ref{sec:SOC}). Then, from this point of view, there is no fundamental difference between RuO$_2$ and other weak ferromagnets.

\par Nevertheless, the weak spin ferromagnetism in rutile compounds is allowed by symmetry~\cite{DzyaloshinskiiRutile} and the microscopic mechanism behind it is believed to be the single-ion anisotropy~\cite{Moriya_NiF2}. However, the single-ion anisotropy appears only at second order in the SO coupling. Therefore, we expect it to play only a minor role in the AHE and in the orbital magnetization, both of which arise already at first order.

\par We assume that the N\'eel field is parallel to $x$. Then, according to Eqs.~(\ref{eq:gbtx}) and (\ref{eq:gbty}), in the local coordinate frame specified by $\boldsymbol{q}=\boldsymbol{G}_{1}$ and $\alpha=0$, the SO interaction is described by the Pauli matrices $\hat{\sigma}_{y}$ and $\hat{\sigma}_{z}$. Other transformations are given by Eqs.~(\ref{eq:gbt1}) and (\ref{eq:gbt2}). 

\par Therefore, we will deal with the following Hamiltonian:  
\noindent
\begin{equation}
\hat{\cal H}_{\boldsymbol{k}} = h^{\phantom{0}}_{\boldsymbol{k}} \hat{\mathbb{1}}  + h^{1}_{\boldsymbol{k}} \hat{\sigma}_{x} - h^{y}_{\boldsymbol{k}}\hat{\sigma}_{y} + \left( B + h^{z}_{\boldsymbol{k}} \right)\hat{\sigma}_{z},
\label{eq:Hkr}
\end{equation} 
\noindent where it is again convenient to shift the $\boldsymbol{k}$-mesh: $\boldsymbol{k} \to \boldsymbol{k} + (\pi,0,0)$. Then, the parameters of Eq.~(\ref{eq:Hkr}) will be given by $h^{\phantom{0}}_{\boldsymbol{k}} = 2(t_{2} - \delta t_{2}) (\cos k_{x} + \cos k_{y}) + 2(t_{2} + \delta t_{2}) \cos k_{z}$, $h^{1}_{\boldsymbol{k}} = 8 t_{1} \cos \frac{k_{x}}{2} \cos \frac{k_{y}}{2} \cos \frac{k_{z}}{2}$, $h^{y}_{\boldsymbol{k}} = -8 t_{x} \sin \frac{k_{x}}{2} \cos \frac{k_{y}}{2} \sin \frac{k_{z}}{2}$, and $h^{z}_{\boldsymbol{k}} = 8 t_{y} \cos \frac{k_{x}}{2} \sin \frac{k_{y}}{2} \cos \frac{k_{z}}{2}$, where $t_{1}$ is the nearest-neighbor hopping between the sublattices and $t_{2} \pm \delta t_{2}$ are the hoppings within the sublattices: along $z$ ($t_{2} + \delta t_{2}$) and in the $xy$ plane ($t_{2} - \delta t_{2}$). Although for the $P4_{2}/mnm$ symmetry $t_{x} = t_{y}$, it is convenient to treat $t_{x}$ and $t_{y}$ as independent SO parameters. The eigenvalues, eigenfunctions, and properties of Hamiltonian (\ref{eq:Hkr}) can be obtained along the same line as in Sec.~\ref{sec:mmodel} after replacing $B$ by $B + h^{z}_{\boldsymbol{k}}$.

\par Furthermore, we have the properties: $h^{\phantom{0}}_{-\boldsymbol{k}} = h^{\phantom{0}}_{\boldsymbol{k}}$, $h^{1}_{-\boldsymbol{k}} = h^{1}_{\boldsymbol{k}}$, $h^{y}_{-\boldsymbol{k}} = h^{y}_{\boldsymbol{k}}$, and $h^{z}_{-\boldsymbol{k}} = -h^{z}_{\boldsymbol{k}}$, which immediately yields: $\varepsilon^{\pm}_{\boldsymbol{k}}(B) = \varepsilon^{\pm}_{-\boldsymbol{k}}(-B)$ and $\varepsilon^{\pm}_{\boldsymbol{k}}(B,t_{y}) = \varepsilon^{\pm}_{\boldsymbol{k}}(-B,-t_{y})$. Then, the Berry curvature satisfies the following properties: $\Omega^{y}_{\boldsymbol{k}}(B) = - \Omega^{y}_{-\boldsymbol{k}}(-B)$ ($\partial_{k_{a}} \phi_{\boldsymbol{k}}$ changes the sign for $a=$ $z$ and $x$, while $\sin 2 \theta_{\boldsymbol{k}}$ and $\partial_{k_{a}} \theta_{\boldsymbol{k}}$ do not change the sign) and $\Omega^{y}_{\boldsymbol{k}}(B,t_{y}) = - \Omega^{y}_{\boldsymbol{k}}(-B,-t_{y})$ ($\partial_{k_{a}} \theta_{\boldsymbol{k}}$ changes the sign, while $\sin 2 \theta_{\boldsymbol{k}}$ and $\partial_{k_{a}} \phi_{\boldsymbol{k}}$ do not change the sign). Therefore, the Hall conductivity, $\sigma_{zx}$, will be odd in $B$, $\sigma_{zx}(-B) = - \sigma_{zx}(B)$, and even in $t_{y}$, $\sigma_{zx}(-t_{y}) = \sigma_{zx}(t_{y})$. Moreover, $\sigma_{zx}$ is odd in $t_{x}$, i.e., when the SO coupling is in the direction of the N\'eel field: $\sigma_{zx}(-t_{x}) = -\sigma_{zx}(t_{x})$. Thus, to calculate $\sigma_{zx}$ to first order in the SO coupling, $t_{y}$ can be neglected. This is consistent with the general strategy for calculating DM interactions using the response theory: in order to calculate some particular (say, $x$) component of the DM vectors, it is sufficient to align the exchange field along $x$ and consider only the $x$ component of the SO coupling~\cite{review2024,Sandratskii}.

\par
\begin{figure}[b]
\begin{center}
\includegraphics[width=8.2cm]{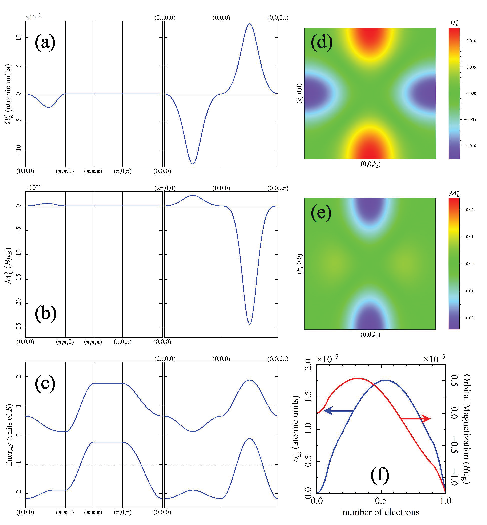}
\end{center}
\caption{Results for RuO$_2$-type systems with the parameters $t_{1}=1$, $t_{2}=1.53$, $\delta t_{2}=1.76$, and $t_{x} = 0.1$ reported in Ref.~\cite{Roig}, and $B/t_{1} = 8$: (a) Berry curvature, (b) orbital magnetization for $n_{\rm el}=0.5$ electrons, and (c) band dispersion along high-symmetry directions of bct Brillouin zone (the Fermi level for $n_{\rm el}=0.5$ is shown by dash-dotted line); Corresponding contour plots of (d) Berry curvature and (c) orbital magnetization in the $k_{y}=0$ plane ($0 \le k_{x} \le 2\pi$ and $0 \le k_{z} \le 2\pi$); (f) Band filling dependence of anomalous Hall conductivity and orbital magnetization.} 
\label{fig:RuO2s}
\end{figure}

 After neglecting $t_{y}$, the Berry curvature can be obtained from Eq.~(\ref{eq:Berryc}) as
\noindent
\begin{equation}
\Omega^{y}_{\boldsymbol{k}} = \frac{4t_{1}t_{x}B}{A^{3}_{\boldsymbol{k}}} \cos^{2} \frac{k_{y}}{2} \left( \cos k_{z} - \cos k_{x} \right).
\label{eq:wykRuO2}
\end{equation}
\noindent The results are summarized in Fig.~\ref{fig:RuO2s}, using parameters reported in Ref.~\cite{Roig}.
\noindent
\noindent The largest contributions to $\Omega^{y}_{\boldsymbol{k}}$ arise from the $(2\pi,0,0) \to (0,0,0) \to (0,0,2\pi)$ directions in the $k_{y}=0$ plane, which are additionally modulated along $k_{y}$, as described by Eq.~(\ref{eq:wykRuO2}). Moreover, $\Omega^{y}_{\boldsymbol{k}}$ has nodal lines along $k_{z}= \pm k_{x}$. Therefore, if the bands along $(0,0,0) \to (2\pi,0,0)$ and $(0,0,0) \to (0,0,2\pi)$ were equally populated, $\sigma_{xy}$ would vanish. That is why we need the ``tetragonal strain'' $\delta t_{2}$~\cite{NakaOrganic,arXiv2025}, which is associated with the tetragonal distortion $c/a<1$ of bct lattice and leads to considerably stronger dispersion along $(0,0,0) \to (0,0,2\pi)$. Then, for the $n_{\rm el}=0.5$ electrons, the ``majority-spin'' band is fully populated along $(0,0,0) \to (2\pi,0,0)$ and only partially populated along $(0,0,0) \to (0,0,2\pi)$, resulting in finite $\sigma_{xy}$. In the expression for the orbital magnetization, Eq.~ (\ref{eq:OMk}), $\Omega^{y}_{\boldsymbol{k}}$ is additionally modulated by $h^{\phantom{0}}_{\boldsymbol{k}}$, thus destroying the perfect cancellation between the directions $(0,0,0) \to (2\pi,0,0)$ and $(0,0,0) \to (0,0,2\pi)$. Unlike in the previous examples, the form of orbital magnetization substantially deviates from the one of anomalous Hall conductivity as is clearly seen in Fig.~\ref{fig:RuO2s}(f). This effect is related to the much stronger tetragonal distortion, $\delta t_{2}/t_{1}=1.76$~\cite{Roig}, when the term $h^{\phantom{0}}_{\boldsymbol{k}} \Omega^{y}_{\boldsymbol{k}}$ in the orbital magnetization starts to play a sizable role. 

\section{\label{sec:separation} Classification based on realistic electronic structure}
\par The existence of AHE without band splitting is relatively new and important finding expected from the analysis of simple one-orbital toy model~\cite{NakaOrganic,arXiv2025}. This model can be justified for a limited number of compounds, like $\kappa$-type organic conductors~\cite{NakaOrganic}, La$_2$CuO$_4$~\cite{arXiv2025}, VF$_4$, and CuF$_2$, where the electronic structure near the Fermi level is indeed featured by a single well-separated band, which can be taken as the basis for the construction of realistic model and using for these purposes the input from first-principles electronic-structure calculations~\cite{review2008,WannierRevModPhys,ImadaPRB,ImadaPRX}. This model is typically in the strong coupling limit meaning that the band splitting, $\varepsilon^{\uparrow}_{\boldsymbol{k}} - \varepsilon^{\downarrow}_{\boldsymbol{k}} = \delta \varepsilon_{\boldsymbol{k}}$, is much weaker than $B$, so that the contribution to the Berry curvature associated with the band splitting is about $\delta \varepsilon_{\boldsymbol{k}}/B$ times smaller than the one coming from degenerate bands~\cite{arXiv2025}.

\par The next important questions is how the contributions related to the band splitting and spin-degeneracy can be separated in realistic electronic structure calculations. For these purposes, one can use results of the model analysis as a guidance. First, it is important to stay in the global coordinate frame: although generalized Bloch theorem is vital for clarifying fundamental aspects of the time-reversal symmetry breaking in centrosymmetric antiferromagnets, it relies on additional approximations, such as the neglect of the spin-splitting of AFM bands, which do not necessarily hold in realistic calculations. Furthermore, the generalized Bloch theorem does not provide a substantial technical advantage for calculating AFM spin structures (in comparison with incommensurate spin spirals). Second, spin should be a good quantum number (otherwise, the analysis of spin-splitting is meaningless). Therefore, one should keep only one component of the SO coupling, which is parallel to the N\'eel field. This is similar to what we did in the model. Two other components of the SO coupling cannot be fully eliminated as in the one-orbital case. However, if the spin canting is small, their contribution to AHE is also expected to be small~\cite{SmejkalSA}. 

\par Suppose that the crystal is oriented in such a way that the N\'eel field and SO coupling are parallel to $z$. Importantly, using this setting and only $z$ component of the SO coupling, we should be able to reproduce $z$ component of DM interactions~\cite{review2024,Sandratskii} and, hopefully, the Hall conductivity $\sigma_{xy}$, which is related to the sign-alternating part of these interactions~\cite{arXiv2025}. The approximation holds to first order in the SO coupling~\cite{review2024,Sandratskii}. Since the Berry curvature is proportional to the SO coupling at lowest order~\cite{NakaOrganic,arXiv2025,Roig}, it should be captured by the present approach.

\par Finally, let us assume that the altermagnetic bands near the Fermi level can be classified by the spin-doublet quantum number $n$ and spin indices, $\uparrow$ or $\downarrow$, describing the splitting of each such doublet. Then, the Hall conductivity, 
\noindent
\begin{displaymath}
\sigma_{xy} =  - \frac{1}{V} \sum_{n} \int_{\rm BZ} \frac{d \boldsymbol{k}}{\Omega_{\rm BZ}} \left( f^{\uparrow}_{n\boldsymbol{k}} \Omega^{\uparrow,z}_{n\boldsymbol{k}} + f^{\downarrow}_{n\boldsymbol{k}} \Omega^{\downarrow,z}_{n\boldsymbol{k}} \right),  
\end{displaymath}
\noindent can be identically rearranged as 
\noindent
\begin{displaymath}
\sigma_{xy} =  - \frac{1}{V} \sum_{n} \int_{\rm BZ} \frac{d \boldsymbol{k}}{\Omega_{\rm BZ}} \left( f^{\rm e}_{n\boldsymbol{k}} \Omega^{{\rm o},z}_{n\boldsymbol{k}} + f^{\rm o}_{n\boldsymbol{k}} \Omega^{{\rm e},z}_{n\boldsymbol{k}} \right),  
\end{displaymath}
\noindent where $\Omega^{{\rm o},z}_{n\boldsymbol{k}} = \frac{1}{2} ( \Omega^{\uparrow,z}_{n\boldsymbol{k}} - \Omega^{\downarrow,z}_{n\boldsymbol{k}} )$ and $f^{\rm o}_{n\boldsymbol{k}} = ( f^{\uparrow}_{n\boldsymbol{k}} - f^{\downarrow}_{n\boldsymbol{k}} )$ are odd in the N\'eel field $B$ (the interchange of $\uparrow$ and $\downarrow$), while $\Omega^{{\rm e},z}_{n\boldsymbol{k}} = \frac{1}{2} ( \Omega^{\uparrow,z}_{n\boldsymbol{k}} + \Omega^{\downarrow,z}_{n\boldsymbol{k}} )$ and $f^{\rm e}_{n\boldsymbol{k}} = ( f^{\uparrow}_{n\boldsymbol{k}} + f^{\downarrow}_{n\boldsymbol{k}} )$ are even. Obviously, total $\sigma_{xy}$ is odd in $B$. Basically, this is a generalization of the one-orbital model~\cite{arXiv2025}. The first term can be finite even when the bands are spin-degenerate, while the second term is finite only when the $\uparrow$ and $\downarrow$ bands are split across the Fermi level, like in the FM state. Indeed, according to Eq.~(\ref{eq:Berryc}), $\Omega^{z}_{n\boldsymbol{k}}$ is odd in $B$ when the bands are spin-degenerate (in the global frame), manifesting the $\{ \mathcal{S} | {\bf t} \}$ symmetry. The even term, $\Omega^{{\rm e},z}_{n\boldsymbol{k}}$, appears when the $\{ \mathcal{S} | {\bf t} \}$ symmetry is broken by the altermagnetic band splitting (see Table~\ref{tab:transformation}). Importantly, although $\Omega^{{\rm o},z}_{n\boldsymbol{k}}$ and $\Omega^{{\rm e},z}_{n\boldsymbol{k}}$ can be viewed as two independent contributions, both of them are induced by the relativistic SO interaction (and the N\'eel field). Therefore, AHE can exist without band splitting. However, it cannot exist without SO interaction, as was discussed in Sec.~\ref{sec:global}.

\par It is instructive to briefly discuss the relative importance of $\Omega^{{\rm o},z}_{n\boldsymbol{k}}$ and $\Omega^{{\rm e},z}_{n\boldsymbol{k}}$ in some characteristic materials using the simplified expression $\Omega^{{\rm e},z}_{n\boldsymbol{k}}/\Omega^{{\rm o},z}_{n\boldsymbol{k}} \sim \delta \varepsilon^{\max}_{n}/B$ anticipated in the strong coupling limit~\cite{arXiv2025}, where $\varepsilon^{\max}_{n}$ is a characteristic (maximal) spin-splitting near the Fermi level (a measure of electron hoppings responsible for this splitting). Particularly, an exceptionally large $\delta \varepsilon^{\max}_{n} = 0.8$ eV has been reported in MnTe on the basis of ARPES measurements~\cite{OghushiMnTe}. However, $B$ in MnTe is also expected to be large. In the local spin density approximation, it can be estimated as $B \sim \frac{1}{2} JM$, where $J$ is a characteristic exchange coupling (typically being of the order of intraatomic Hund's rule coupling, $J \sim 0.8$ eV in the case of Mn$^{2+}$~\cite{PRB1994}) and $M$ is the local spin magnetic moment ($M \sim 4.25 \mu_{\rm B}$ in MnTe~\cite{MnTeDFT}). Therefore, the ratio $\delta \varepsilon^{\max}_{n}/B$ can be estimated as $0.47$. The effects of on-site Coulomb repulsion $U$ are also anticipated~\cite{MnTeDFT,MnTePRL}. Particularly, like in one-orbital model, $U$ can contribute to $B$ and further decrease the ratio $\delta \varepsilon^{\max}_{n}/B$.

\par RuO$_2$ is an exceptional case. According to theoretical calculations~\cite{SmejkalSA}, $\varepsilon^{\max}_{n}$ is large ($\sim 1$ eV), while $M$ is small ($\sim 1.17 \mu_{\rm B}$). Furthermore, $J$ for $4d$ materials is also expected to be relatively small ($\sim 0.6$ eV~\cite{PRB1994}). Then, $\delta \varepsilon^{\max}_{n}/B$ can be estimated as $2.85$. This value will be somewhat decreased by considering the effects of on-site $U$, which are also anticipated in RuO$_2$~\cite{SmejkalSA}. In any case, the even-order contribution $\Omega^{{\rm e},z}_{n\boldsymbol{k}}$ is expected to be large in RuO$_2$. However, it does not necessarily mean that the odd-order contribution $\Omega^{{\rm o},z}_{n\boldsymbol{k}}$ is negligible. Furthermore, whether $\Omega^{{\rm e},z}_{n\boldsymbol{k}}$ will dominate in the calculations of $\sigma_{xy}$ depends on how the topology of $\Omega^{{\rm e},z}_{n\boldsymbol{k}}$ matches the one of the Fermi surface and whether the areas of large spin-splitting of bands across the Fermi level will coincide with the ones where $\Omega^{{\rm e},z}_{n\boldsymbol{k}}$ is large~\cite{arXiv2025}. 

\section{\label{sec:Summary} Summary and Outlook}
\par The time-reversal symmetry breaking in AFM substances implies certain analogies with ferromagnetism. On the one hand, it can lift the Kramers degeneracy and splits the AFM bands with opposite spins. This is an overt analogy, which relies on more or less standard picture of antiferromagnetism where there are two magnetic sublattices with the moments pointing in opposite directions. Today, this analogy is intensively explored in the context of altermagnetism~\cite{SmejkalPRX1,SmejkalPRX2,MazinPRX1}.

\par In the present work, we have argued that there is a deeper analogy, which allows us to present certain classes of antiferromagnets with broken $\mathcal{T}$ as if they were ferromagnets with only one magnetic site per cell. 

\par Such a representation appears to be possible due to the hidden $\{ \mathcal{S} | {\bf t} \}$ symmetry of SO interaction in centrosymmetric antiferromagnets, allowing transformation to the local coordinate frame where all the spins are pointed along the positive direction of $z$, i.e. like in ferromagnets. This symmetry is the consequence of the inversional invariance of the system. The key point is that the inversion symmetry can be broken locally, in individual bonds. Nevertheless, in centrosymmetric antiferromagnets, some inversion centers continue to exist, so that $\mathcal{I}$ transforms the AFM lattice to itself. If the system is subjected to the lattice distortion, in order to preserve $\mathcal{I}$, this distortion must be antipolar. In the other words, we are dealing with the \emph{antipolar antiferromagnetism}.

\par The antipolar distortion inevitably leads to the doubling of the crystallographic cell, where the SO interaction behaves as an antiferromagnetically ordered object and changes its sign when one moves from one magnetic sublattice to another. Nevertheless, this sign change of the SO interaction can be compensated in some local coordinate frame. 

\par Indeed, the $\{ \mathcal{S} | {\bf t} \}$ symmetry allows us to use the generalized Bloch theorem, which greatly simplify the first-principles calculations of incommensurate spin-spiral states~\cite{Sandratskii_review,Mryasov}. Using generalized Bloch theorem, any such spiral with an arbitrary propagation vector $\boldsymbol{q}$ can be described within one crystallographic unit cell. The AFM alignment of spins is a particular case of the spin spiral. Nevertheless, the generalized Bloch theorem is typically at odds with the relativistic SO interaction, which tends to lock the spins along one particular direction and, thus, act against the spin spiral~\cite{Sandratskii}. The antiferromagnetism appears to be a special case: even though the generalized Bloch theorem is not valid for an arbitrary propagation $\boldsymbol{q}$, it can still be valid for the AFM alignment in antipolar lattice, which imposes a constrain on the form of the SO interaction. As the result, in the local coordinate frame, the N\'eel field \emph{and} SO interaction become translationally invariant on the ``ferromagnetic'' lattice with one magnetic site per cell. 

\par The regular AFM order doubles the unit cell. In unconventional antiferromagnets with broken $\mathcal{T}$, the magnetic unit cell should be as large as the crystallographic one~\cite{Dzyaloshinskii1991}. We substantially revise this canonical point of view and show that, due to the $\{ \mathcal{S} | {\bf t} \}$ symmetry, the magnetic unit cell of centrosymmetric antiferromagnets can be even \emph{smaller} than the crystallographic one. The situation is highly unusual, but reflects the fundamental symmetry of these systems and naturally explains their similarity with ferromagnets. The emergence of AHE, orbital magnetization, and magneto-optical effect in this case becomes natural and all these ``ferromagnetic'' properties can be evaluated in the same way as for regular ferromagnets, but in the local coordinate frame. 

\par Because the $\{ \mathcal{S} | {\bf t} \}$ symmetry protects the AFM bands degeneracy, this case is qualitatively distinct from altermagnetism. These phenomena are related to different symmetries, which typically coexist in one magnetic space group: the $\{ \mathcal{S} | {\bf t} \}$ symmetry is a consequence of inversional invariance, while the band splitting occurs due to rotational, typically $\{ \mathcal{C}|{\bf t} \}$, symmetry, which is combined with $\mathcal{T}$. $\{ \mathcal{C}|{\bf t} \}$ does not conflict with the $\{ \mathcal{S} | {\bf t} \}$ symmetry of the SO interaction, which operates between different sublattices. The main effect of $\{ \mathcal{C}|{\bf t} \}$ on the SO interaction is that it relates the parameters of this interaction on the symmetry-related bonds around each magnetic site. However, $\{ \mathcal{C}|{\bf t} \}$ also specifies the form of hopping parameters operating in different sublattices, which are related by the rotation $\mathcal{C}$. This can split the AFM bands and eventually break the $\{ \mathcal{S} | {\bf t} \}$ symmetry. 

\par If microscopic Hamiltonian were real, the transformation $\{ \mathcal{S} | {\bf t} \}$ would be identical to $\{ \mathcal{T} | {\bf t} \}$. However, 
when the bonds connecting different sublattices are noncentrosymmetric, as in antipolar lattices, the Hamiltonian is complex. Then, $\{ \mathcal{S} | {\bf t} \}$ is not the same as $\{ \mathcal{T} | {\bf t} \}$. Therefore, the time-reversal symmetry can be broken, but the Hamiltonian remain invariant under $\{ \mathcal{S} | {\bf t} \}$, keeping the bands degenerate. This is another fundamental reason why the centrosymmetric antiferromagnets can exhibit AHE and other ferromagnetic phenomena even if the bands are spin degenerate. The fundamental $\{ \mathcal{S} | {\bf t} \}$ symmetry establishes the connection between unconventional antiferromagnetism with spin-degenerate bands and conventional ferromagnetism with spin-split bands.

\par The basic idea was illustrated on a number of examples, including square perovskite lattice, monoclinic VF$_4$ and CuF$_2$, and antiferromagnetism in the rutile compounds. Nevertheless, it has much wider implication to the properties of unconventional antiferromagnets. For instance, the authors of recent review article~\cite{LingBai} have proposed 221 three-dimensional materials, which could potentially be altermagnetic, and this number continues to grow. However, most of these materials are centrosymmetric antiferromagnets, whose fundamental ferromagnetic properties are consequences of the hidden $\{ \mathcal{S} | {\bf t} \}$ symmetry.

\par The band splitting provides an additional contribution to AHE, which is also related to relativistic SO interaction. However, in model considerations, it emerges only as a small correction to AHE obtained for the spin-degenerate bands~\cite{arXiv2025}. Moreover, for certain materials, like La$_2$CuO$_4$, the band splitting vanishes due to the special symmetry constraint imposed on parameters of the one-orbital model, which is expected to provide an adequate description near the Fermi level~\cite{ImadaPRB,ImadaPRX}.

\par In the light of these arguments, the systematic experimental information establishing a correlation between the band splitting and AHE is highly welcomed. Today, this information is very sparse. For instance, AHE has been experimentally measured in MnTe~\cite{MnTePRL} and attributed to the large band splitting~\cite{OghushiMnTe,BelashchenkoMnTe,MnTePRL}. AHE has been also reported in AFM FeS~\cite{FeS}, but no information about the band splitting is currently available. As an opposite example, no indications of the magnon band splitting has been reported in MnF$_2$~\cite{MoranoMnF2}, which is yet another altermagnetic candidate crystallizing in the rutile structure~\cite{LingBai}. Although this is not a direct probe of the band splitting, some relationship is anticipated between them because interatomic exchange interactions should reflect details of the electronic structure~\cite{review2024}. Our work clearly shows that MnF$_2$ can exhibit the magneto-optical effect or AHE (for instance, under the doping) irrespectively on whether there is a band splitting or not. An interesting aspect of MnF$_2$ is that the $d^{5}$ configuration of Mn$^{2+}$ is nearly spherical. Therefore, the alternation of exchange interactions (and hopping parameters) responsible for the band splitting is expected to be small, so as the single-ion anisotropy. In this sense, the finding of Ref.~\cite{MoranoMnF2} is quite natural. Although in the ground state the magnetic moments in MnF$_2$ are aligned parallel to $z$, that prohibits the magneto-optical activity, they can be rotated into the basal plane via the spin-flop transition in magnetic field~\cite{FelcherKleb}. 

\par The spin group theory is a popular tool for identifying possible band splitting in altermagnetic materials~\cite{SmejkalPRX1,ChenPRX}. Nevertheless, the symmetry alone does not say whether this splitting is large or small. It can be accidentally zero (as in MnF$_2$) or identically zero (as in the one-orbital model of La$_2$CuO$_4$). On the other hand, the SO interaction, which leads to AHE in antipolar antiferromagnets, is always finite due to the lack of bond inversion symmetry. Formally, we deal with the same materials, which were recently identified as the novel Type-II phase of magnetism (altermagnetism)~\cite{SmejkalPRX1}. However, these materials have a long history~\cite{Dzyaloshinskii_weakF,DzyaloshinskiiPM} and were known before as centrosymmetric antiferromagnets~\cite{TurovBook,TurovUFN}. The centrosymmetric antiferromagnetism is a more general definition, implying a wide range of ``ferromagnetic'' properties, which are not necessarily related to the spin splitting of bands. The altermagnetism (the band splitting) is just a subclass of centrosymmetric antiferromagnetism. 

\par The lack of inversion symmetry would break the $\{ \mathcal{S} | {\bf t} \}$ symmetry of SO interaction. For instance, in the simplest polar magnets, the SO interaction is translationally invariant and transformed to itself by ${\bf t}$. This induces DM interactions connecting different unit cells, which are responsible for the formation of noncollinear magnetic textures, such as spin spirals or skyrmions~\cite{Dzyaloshinskii1964,bog1,bog2}. Thus, generally such systems cannot be classified as ``collinear antiferromagnets''. Nevertheless, the lack of inversion symmetry can lead to a number of interesting properties, including skyrmion-induced topological Hall effect~\cite{NagaosaTokura}.  

\par Another prominent example of topological magnetism is the noncoplanar antiferromagnetism realized in triangular, pyrochlore, or kagome lattices~\cite{Ohgushi,Nakatsuji,FengNC,WatanabePRB}. The main difference from nearly collinear centrosymmetric antiferromagnets considered in the present work is that in noncoplanar magnets, AHE is induced by scalar spin chirality without SO interaction. 

\par Thus, the altermagnetic splitting of bands is certainly an interesting theoretical discovery leading to a number of practically important phenomena such as the spin-current generation in AFM substances~\cite{Naka,JunweiLiuNC,Naka_Spintronics}. However, regarding the whole spectrum of properties expected in centrosymmetric antiferromagnets with broken time-reversal symmetry, it would not be right to attribute all of them to the band splitting and try to consider all of them only from the viewpoint of this splitting. Even if the band splitting is small, the material is still expected to host robust AHE and net orbital magnetization, exhibit magneto-optical rotations and other ``ferromagnetic'' phenomena, which are related to other, hidden, symmetries of such AFM state, being the consequence of the inversional invariance. 

\section*{Acknowledgement}
\par I am indebted to Sergey Nikolaev and Akihiro Tanaka for collaboration on earlier stages of this project~\cite{arXiv2025}. MANA is supported by World Premier International Research Center Initiative (WPI), MEXT, Japan.

\

\par The author declares no competing interests.

\appendix
\section{\label{sec:SOC} Elimination of weakly ferromagnetic components of the spin-orbit coupling}
\par To be specific, let us consider the case where $x$ and $z$ components of the SO coupling between the sublattices $1$ and $2$ have the same sign in all neighboring bonds (and responsible for weak spin ferromagnetism), while the $y$ component is sign-alternating (and responsible for AHE), which is rather common for two-sublattice antiferromagnets~\cite{arXiv2025}. The complex transfer integrals, including the SO coupling, are given by $t_{\boldsymbol{R},\boldsymbol{R}'} \hat{\mathbb{1}} + i\boldsymbol{t}_{\boldsymbol{R},\boldsymbol{R}'} \cdot \hat{\boldsymbol{\sigma}}$. 

\par Then, let us consider the unitary transformation of the sublattice $2$, $\hat{U} = e^{i\varphi \boldsymbol{n} \cdot \hat{\boldsymbol{\sigma}}} = \cos \frac{\varphi}{2}\hat{\mathbb{1}} + i \boldsymbol{n} \cdot \hat{\boldsymbol{\sigma}} \sin \frac{\varphi}{2}$, such that $\left( t_{\boldsymbol{R},\boldsymbol{R}'} \hat{\mathbb{1}} + i\boldsymbol{t}_{\boldsymbol{R},\boldsymbol{R}'} \cdot \hat{\boldsymbol{\sigma}} \right) \hat{U} \approx \tilde{t}_{\boldsymbol{R},\boldsymbol{R}'} \hat{\mathbb{1}} + i t^{y}_{\boldsymbol{R},\boldsymbol{R}'} \hat{\sigma}_{y}$. In principle, by using such transformation, one can completely eliminate the SO term $i\boldsymbol{t}_{\boldsymbol{R},\boldsymbol{R}'} \cdot \hat{\boldsymbol{\sigma}}$ \emph{separately} in each bond~\cite{Shekhtman,Kaplan}. However, this cannot be done \emph{simultaneously} for all the bonds, unless the parameters of the SO coupling are the same in all these bonds. Nevertheless, one can try to eliminate $t^{x}_{\boldsymbol{R},\boldsymbol{R}'} \equiv t_{x}$ and $t^{z}_{\boldsymbol{R},\boldsymbol{R}'} \equiv t_{z}$, which are the same in all the bonds. 

\par Assuming $\boldsymbol{n} = \frac{(t_{x},0,t_{z})}{\sqrt{t_{x}^2+t_{z}^2}}$, which does not depend on the alternating $y$-component, one can find the following equations for $\varphi$:
\noindent
\begin{displaymath}
t_{x} \cos \frac{\varphi}{2} + n_{x}t \sin \frac{\varphi}{2} - [\boldsymbol{t}_{\boldsymbol{R},\boldsymbol{R}'} \times \boldsymbol{n}]_{x} \sin \frac{\varphi}{2} = 0
\end{displaymath}
\noindent and 
\noindent
\begin{displaymath}
t_{z} \cos \frac{\varphi}{2} + n_{z}t \sin \frac{\varphi}{2} - [\boldsymbol{t}_{\boldsymbol{R},\boldsymbol{R}'} \times \boldsymbol{n}]_{z} \sin \frac{\varphi}{2} = 0 ,
\end{displaymath}
\noindent where $[\boldsymbol{t}_{\boldsymbol{R},\boldsymbol{R}'} \times \boldsymbol{n}]_{x} = t_{\boldsymbol{R},\boldsymbol{R}'}^{y} n_{z}$ and $[\boldsymbol{t}_{\boldsymbol{R},\boldsymbol{R}'} \times \boldsymbol{n}]_{z} = -t_{\boldsymbol{R},\boldsymbol{R}'}^{y} n_{x}$.
\noindent The first two terms in these equations are first order in the SO coupling, whereas the third term is second order and can be neglected (note that $\varphi$ is first order). Therefore, one can find that $\varphi = -2 \arctan \frac{\sqrt{t_{x}^2+t_{z}^2}}{t}$.

\section{\label{sec:CuF2model} Electronic structure and minimal model for VF$_4$ and CuF$_2$}
\par
\begin{figure}[t]
\begin{center}
\includegraphics[width=4.2cm]{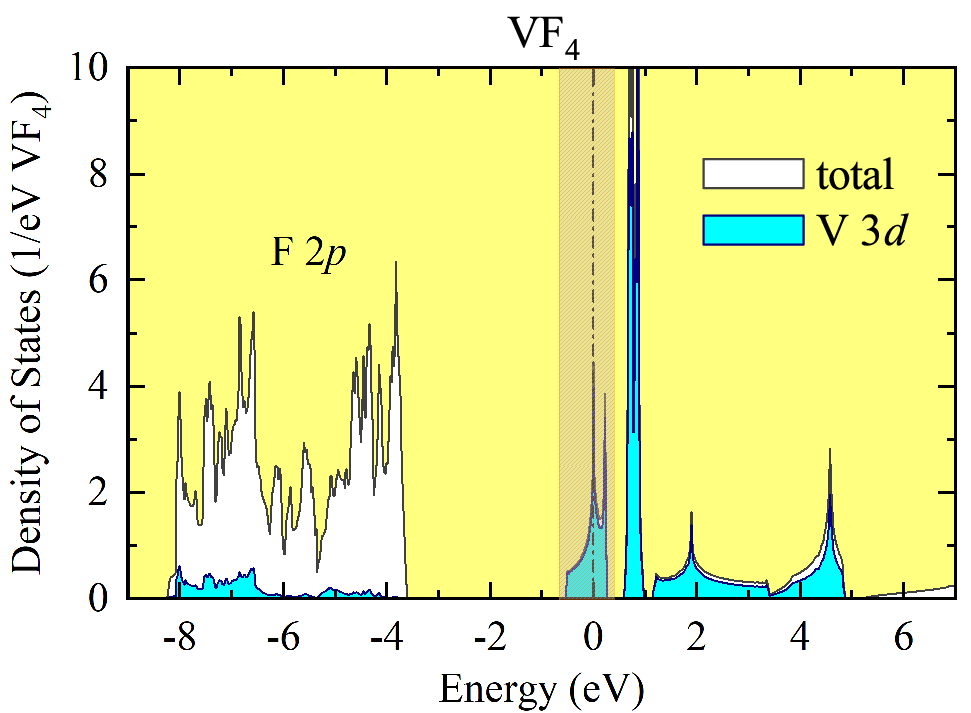} \includegraphics[width=4.2cm]{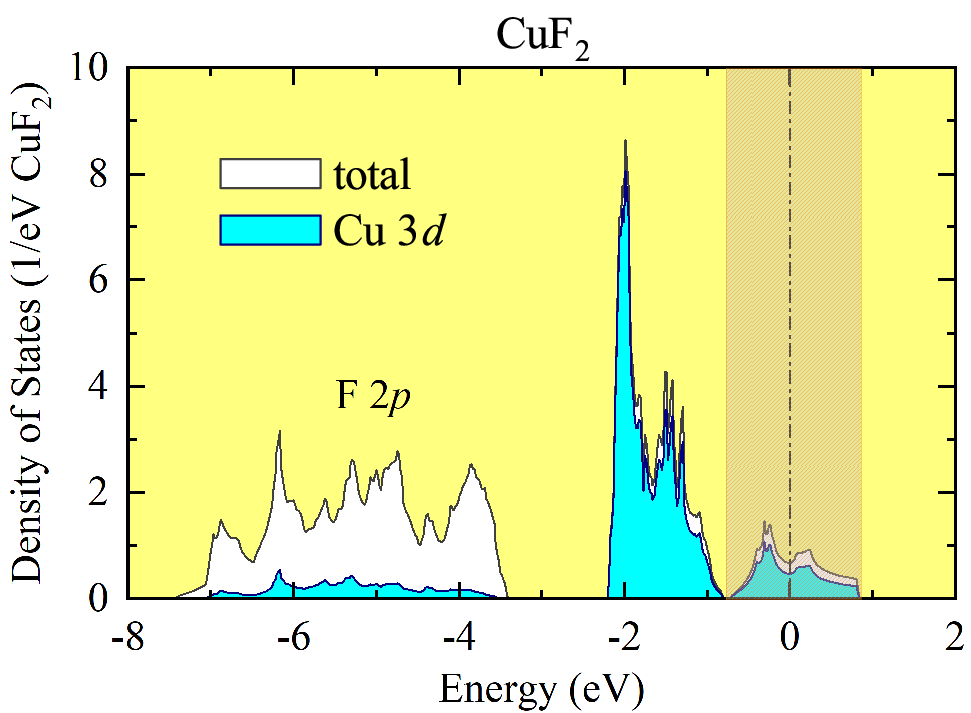} 
\end{center}
\caption{Electronic structure of VF$_4$ (left) and CuF$_2$ (right) in LDA. The Fermi level is at zero energy. Shaded areas depict the bands, which were used for the construction of minimal one-orbital models.} 
\label{fig:CuF2}
\end{figure}

 The electronic structure of VF$_4$ and CuF$_2$ in LDA with the experimental lattice parameters~\cite{VF4,CuF2} is shown in Fig.~\ref{fig:CuF2}.
\noindent
\noindent The calculations are performed using the linear muffin-tin orbital (LMTO) method~\cite{LMTO1,LMTO2} and Vosko-Wilk-Nusair parametrization for the exchange-correlation potential in LDA~\cite{VWN}.
  
\par In both cases, the electronic structure is featured by two well isolated bands at the Fermi level, which are depicted in Fig.~\ref{fig:CuF2}. For each spin, these bands can accommodate one electron. Therefore, in non-magnetic LDA, the bands are half-filled. 

\par Using Bloch functions for these bands, one can construct the localized Wannier functions, which serve as the basis for a minimal model. For these purposes we employ the projector-operator technique~\cite{review2008,WannierRevModPhys}. Then, the transfer integrals $t_{\boldsymbol{R},\boldsymbol{R}'} \hat{\mathbb{1}} + i\boldsymbol{t}_{\boldsymbol{R},\boldsymbol{R}'} \cdot \hat{\boldsymbol{\sigma}}$ can be identifies with the matrix elements of LDA Hamiltonian with the SO coupling in the Wannier basis. The screened on-site Coulomb repulsion $U$ can be evaluated within constrained random-phase approximation~\cite{cRPA}, as explained in Ref.~\cite{review2008}. At the half-filling, the N\'eel field is related to $U$ as $B \approx U/2$.

\par
\begin{figure}[b]
\begin{center}
\includegraphics[width=4.2cm]{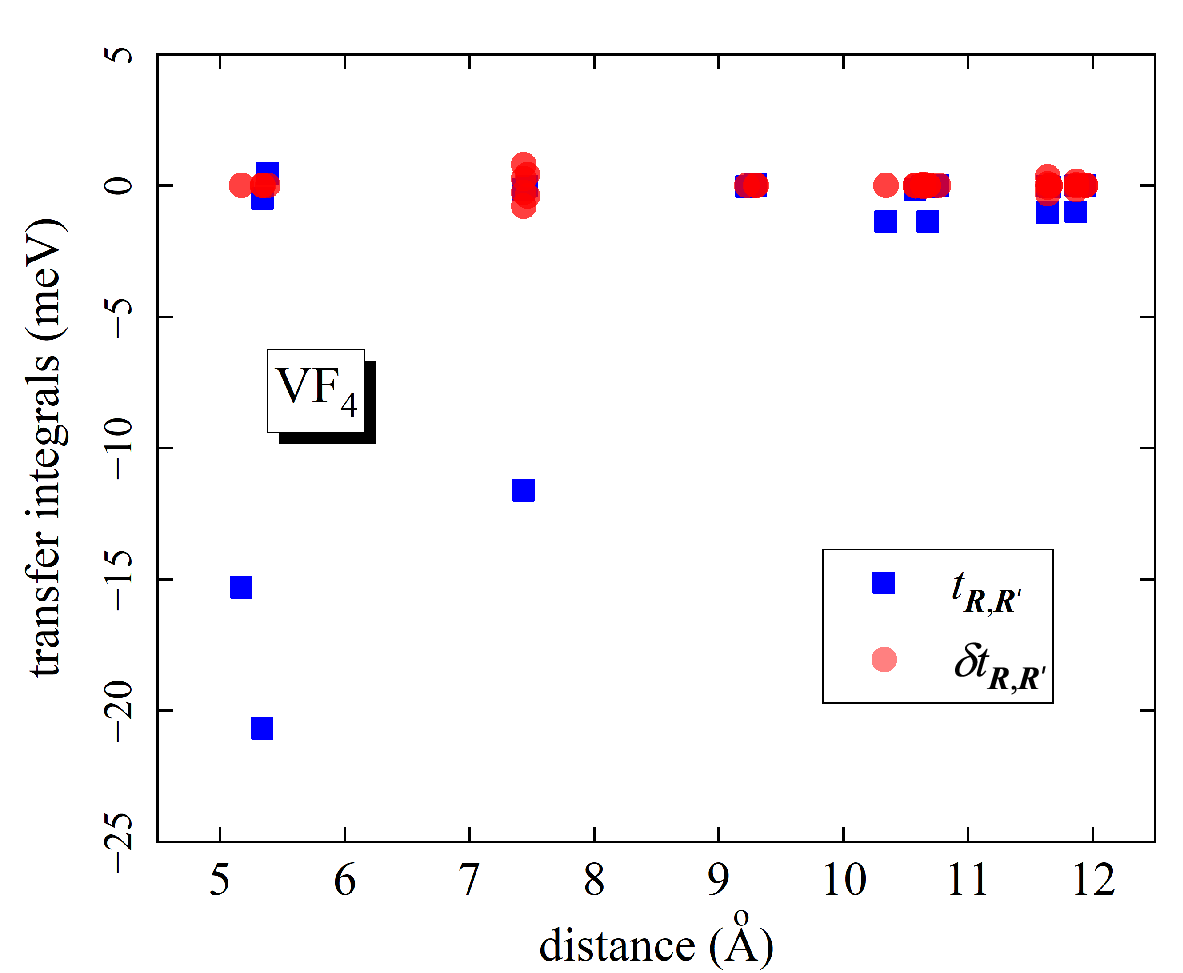} \includegraphics[width=4.2cm]{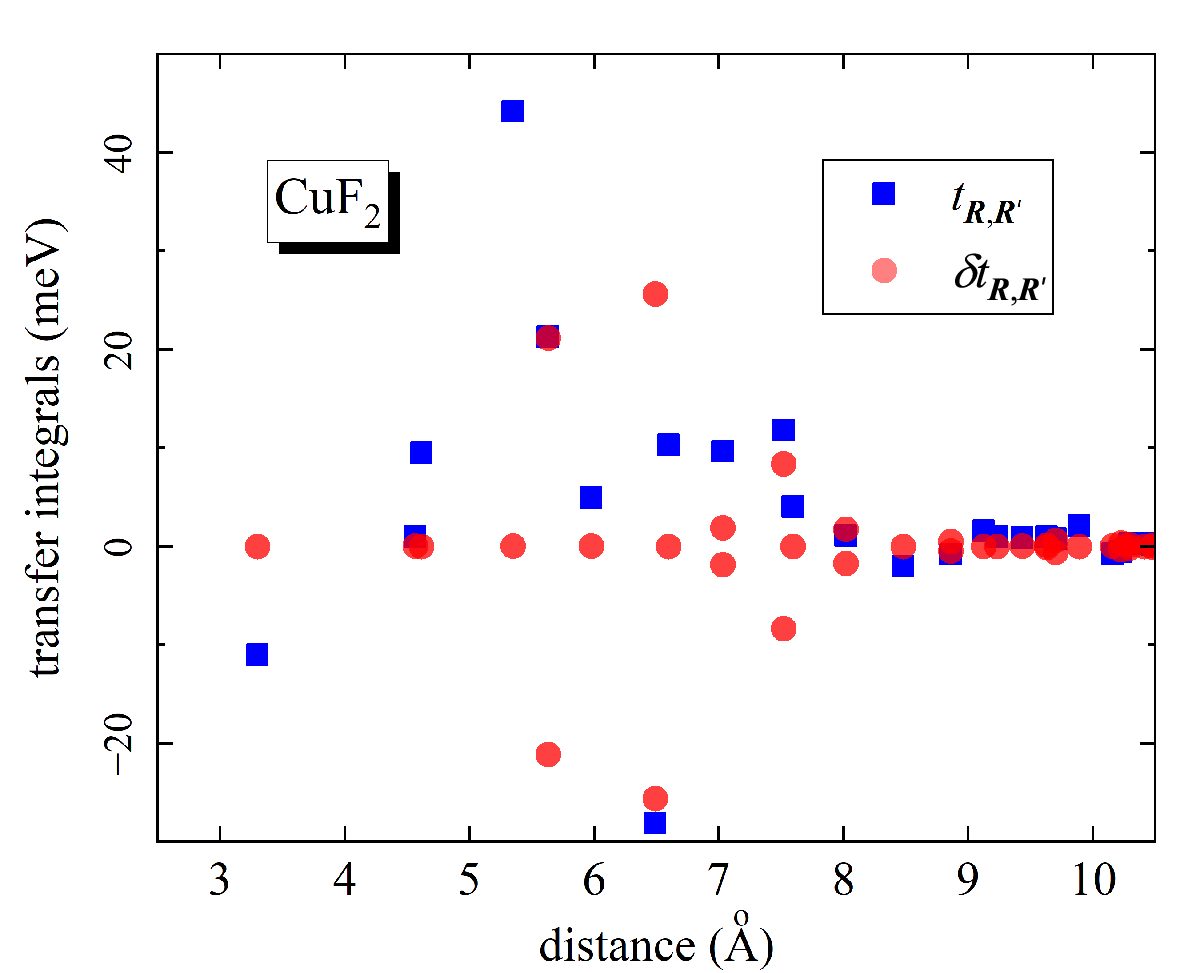} 
\end{center}
\caption{Distance-dependence of regular ($t^{\phantom{I}}_{\boldsymbol{R},\boldsymbol{R}'}$) and altermagnetic ($\delta t^{\phantom{I}}_{\boldsymbol{R},\boldsymbol{R}'}$) transfer integrals operating within the same sublattices in VF$_4$ and CuF$_2$.} 
\label{fig:talter}
\end{figure}

 The transfer integrals operating within the same sublattices were additionally averaged. Namely, using the transfer integrals in the sublattices $I$ and $II$, one can define $t^{\phantom{I}}_{\boldsymbol{R},\boldsymbol{R}'} = \frac{1}{2}(t^{I}_{\boldsymbol{R},\boldsymbol{R}'} + t^{II}_{\boldsymbol{R},\boldsymbol{R}'})$ (the regular contribution) and $\delta t^{\phantom{I}}_{\boldsymbol{R},\boldsymbol{R}'} = \frac{1}{2}(t^{I}_{\boldsymbol{R},\boldsymbol{R}'} - t^{II}_{\boldsymbol{R},\boldsymbol{R}'})$ (the altermagnetic contribution). The approximation consists in neglecting $\delta t^{\phantom{I}}_{\boldsymbol{R},\boldsymbol{R}'}$. This suppress the spin-splitting of the AFM bands, but allows us to employ the generalized Bloch theorem and describe these AFM systems as they would have only one magnetic site in the unit cell. The validity of this approximation depends on the system: while $\delta t^{\phantom{I}}_{\boldsymbol{R},\boldsymbol{R}'}$ and $t^{\phantom{I}}_{\boldsymbol{R},\boldsymbol{R}'}$ are comparable in CuF$_2$, $\delta t^{\phantom{I}}_{\boldsymbol{R},\boldsymbol{R}'}$ appear to be much smaller than $t^{\phantom{I}}_{\boldsymbol{R},\boldsymbol{R}'}$ in VF$_4$ (see Fig.~\ref{fig:talter}).
\noindent


\begin{thebibliography}{99}
\bibitem{footnote}
Strictly speaking, we consider a specific situation when the magnetic sites are located in the inversion centers. Another possibility is when the magnetic site itself is not in the centrosymmetric position, but $\mathcal{I}$ connects two magnetic sites~\cite{TurovBook}. Such a situation is realized, for instance, in Fe$_2$O$_3$~\cite{Dzyaloshinskii_weakF}. Nevertheless, we note that in the latter case, one can treat the magnetic molecule, consisting of two Fe atoms across the inversion center, as a formal lattice site. Then, the situation will be reduced to the one considered in the present work.
\end{thebibliography}
\end{document}